\documentclass[10pt,article]{IEEEtran}

\usepackage[pdftex]{graphicx}
\usepackage{generic}
\usepackage{hyperref}
\usepackage{booktabs}
\usepackage{multirow}
\usepackage[table,xcdraw]{xcolor}
\usepackage{subfigure}
\newsavebox{\imagebox}
\usepackage{array}
\renewcommand{\arraystretch}{1.3}

\usepackage{amsmath}
\usepackage{comment}
\usepackage[utf8]{inputenc}
\usepackage[]{lineno}
\nolinenumbers

\def\eg{\textit{e.g.}}
\def\ie{\textit{i.e.}}

\def\etc{\textit{etc}}

\begin{document}

\title{Automated Parkinson's Disease Detection and Affective Analysis from  Emotional EEG Signals}
%
%
% author names and IEEE memberships
% note positions of commas and nonbreaking spaces ( ~ ) LaTeX will not break
% a structure at a ~ so this keeps an author's name from being broken across
% two lines.
% use \thanks{} to gain access to the first footnote area
% a separate \thanks must be used for each paragraph as LaTeX2e's \thanks
% was not built to handle multiple paragraphs
%
%
%\IEEEcompsocitemizethanks is a special \thanks that produces the bulleted
% lists the Computer Society journals use for "first footnote" author
% affiliations. Use \IEEEcompsocthanksitem which works much like \item
% for each affiliation group. When not in compsoc mode,
% \IEEEcompsocitemizethanks becomes like \thanks and
% \IEEEcompsocthanksitem becomes a line break with idention. This
% facilitates dual compilation, although admittedly the differences in the
% desired content of \author between the different types of papers makes a
% one-size-fits-all approach a daunting prospect. For instance, compsoc 
% journal papers have the author affiliations above the "Manuscript
% received ..."  text while in non-compsoc journals this is reversed. Sigh.

\author{Ravikiran~Parameshwara,~\IEEEmembership{Student~Member,~IEEE},~Soujanya~Narayana,~\IEEEmembership{Student~Member,~IEEE}, ~Murugappan~Murugappan,~\IEEEmembership{~Senior~ Member,~IEEE}, Ramanathan~Subramanian,~\IEEEmembership{Senior~Member,~IEEE}, Ibrahim~Radwan,~\IEEEmembership{Member,~IEEE}, and~Roland~Goecke,\IEEEmembership{~Senior~Member,~IEEE}
\thanks{Ravikiran Parameshwara, Soujanya Narayana, Ramanathan Subramanian, Ibrahim Radwan, and Roland Goecke are with the Human-Centred Technology Research Centre, Faculty of Science and Technology, University of Canberra, ACT, Australia. Murugappan Murugappan is with the Kuwait College of Science and Technology, Kuwait.}}% <-
\markboth{Draft Version}%
{Draft Version}
\maketitle
\begin{abstract}
While Parkinson's disease (PD) is typically characterized by motor disorder, there is evidence of diminished emotion perception in PD patients. This study examines the utility of affective Electroencephalography (EEG) signals to understand emotional differences between PD vs Healthy Controls (HC), and for automated PD detection. Employing traditional machine learning and deep learning methods, we explore (a) dimensional and categorical emotion recognition, and (b) PD vs HC classification from emotional EEG signals. Our results reveal that PD patients comprehend arousal better than valence, and amongst emotion categories, \textit{fear}, \textit{disgust} and  \textit{surprise} less accurately, and \textit{sadness} most accurately. Mislabeling analyses confirm confounds among opposite-valence emotions with PD data. Emotional EEG responses also achieve near-perfect PD vs HC recognition. {Cumulatively, our study demonstrates that (a) examining \textit{implicit} responses alone enables (i) discovery of valence-related impairments in PD patients, and (ii) differentiation of PD from HC, and (b) emotional EEG analysis is an ecologically-valid, effective, facile and sustainable tool for PD diagnosis vis-\'a-vis self reports, expert assessments and resting-state analysis.}  
\end{abstract}

% Note that keywords are not normally used for peerreview papers.
%}

% make the title area
%\maketitle

% To allow for easy dual compilation without having to reenter the
% abstract/keywords data, the \IEEEtitleabstractindextext text will
% not be used in maketitle, but will appear (i.e., to be "transported")
% here as \IEEEdisplaynontitleabstractindextext when the compsoc 
% or transmag modes are not selected <OR> if conference mode is selected 
% - because all conference papers position the abstract like regular
% papers do.
%\IEEEdisplaynontitleabstractindextext
% \IEEEdisplaynontitleabstractindextext has no effect when using
% compsoc or transmag under a non-conference mode.

% For peer review papers, you can put extra information on the cover
% page as needed:
% \ifCLASSOPTIONpeerreview
% \begin{center} \bfseries EDICS Category: 3-BBND \end{center}
% \fi
%
% For peerreview papers, this IEEEtran command inserts a page break and
% creates the second title. It will be ignored for other modes.
\IEEEpeerreviewmaketitle

\section{Introduction}\label{sec:introduction}
% Computer Society journal (but not conference!) papers do something unusual
% with the very first section heading (almost always called "Introduction").
% They place it ABOVE the main text! IEEEtran.cls does not automatically do
% this for you, but you can achieve this effect with the provided
% \IEEEraisesectionheading{} command. Note the need to keep any \label that
% is to refer to the section immediately after \section in the above as
% \IEEEraisesectionheading puts \section within a raised box.

% The very first letter is a 2 line initial drop letter followed
% by the rest of the first word in caps (small caps for compsoc).
% 
% form to use if the first word consists of a single letter:
% \IEEEPARstart{A}{demo} file is ....
% 
% form to use if you need the single drop letter followed by
% normal text (unknown if ever used by the IEEE):
% \IEEEPARstart{A}{}demo file is ....
% 
% Some journals put the first two words in caps:
% \IEEEPARstart{T}{his demo} file is ....
% 
% Here we have the typical use of a "T" for an initial drop letter
% and "HIS" in caps to complete the first word.

\IEEEPARstart{P}{arkinson's} disease (PD) is a neurodegenerative disorder of the central nervous system that affects movements, often causing tremors. PD is characterised by the progressive loss of dopaminergic neurons in the \emph{substantia nigra}~\cite{dauer2003parkinson}. Beside motor dysfunctions, cognitive, behavioural and emotional defects are common in PD~\cite{papagno2018cognitive, trojano2018cognitive}, affecting over 10 million people globally~(\emph{\url{https://www.parkinson.org/}}).

{A number of studies have examined motor and cognitive impairments in PD patients by examining \textit{explicit} user responses (\eg, performance in recognition tasks, self-reports) or \textit{implicit} responses such as Electroencephalogram (EEG) signals~\cite{Wang2020}. Some works detect PD from abnormalities in {resting-state} EEG~\cite{Han2013,Chat17,Shu2020}. Resting-state EEG is acquired in a highly controlled setting, \eg, requiring the subject to remain motionless with eyes closed in a dim and quiet room, which makes this setting {ecologically invalid}. A more realistic setting involves EEG acquisition during routine tasks such as music listening~\cite{EEG_music} or movie watching~\cite{yuvaraj2016brain}}.

{As movie and musical stimuli are} {often emotion eliciting~\cite{DECAF}, {they enable researchers to understand how PD patients perceive emotions. Prior research has identified emotion deficits in PD patients, as they emote less spontaneously to emotion eliciting video clips \cite{smith1996spontaneous} and unpleasant odours \cite{simons2003disturbance}, and their posed facial expressions are disturbed and impaired \cite{jacobs1995emotional}}. Apart from specific emotions, a few studies focus on PD perception of 
the \emph{\textbf{valence}} (feeling of \emph{pleasantness/aversion}) and \emph{\textbf{arousal}} (emotional \emph{intensity}) dimensions~\cite{russell1980circumplex}. Prior studies show that PD patients have a deficit in recognising positive and negative valence emotions from prosody \cite{paulmann2010dynamic} and facial expressions \cite{clark2008specific}, and reduced reactivity to highly arousing pictures \cite{miller2009startle}. Recognising emotions is critical to successful social interaction and communication, apart from inferring non-verbal social behavior such as emotional voice and facial expressions \cite{blair2003facial}. 

%Although some studies observe reduced emotion perception in PD patients, others find that their emotion recognition capabilities are intact~\cite{adolphs1998intact}.
Implicit physiological or biosignals reflect characteristic activity of the central nervous system, and cannot be intentionally suppressed. Recent studies have extensively employed biosignals~\cite{valenza2011role,Ascertain,Shukla2020} for emotion perception in healthy subjects. EEG, functional Magnetic Resonance Imaging (fMRI), Magnetoencephalogram (MEG) and Positron Emission Tomography (PET) provide reliable information on emotional states compared to other modalities \cite{petrantonakis2010emotion}. EEG is non-invasive, has high temporal resolution and can detect changes in brain activity over a span of milliseconds. EEG frequency bands are known to correlate with emotions~\cite{balconi2008consciousness, aftanas2006neurophysiological}. Handcoded EEG descriptors such as Spectral Power Vectors enable emotion detection while Convolutional Neural Networks (CNNs) can automatically learn cognitive and emotional correlates~\cite{bashivan2015learning,Shukla2020}. 
% and EEG signals exhibit non-linear behaviour, requiring non-linear analysis for modeling signal dynamics~\cite{stam1999dynamics}
%

%, employing traditional machine learning and deep learning techniques

This study examines EEG-based PD emotion perception via a comparative analysis of data acquired from PD patients vis-\'a-vis Healthy Controls (HC). We explore both low-level EEG descriptors such as Spectral Power Vectors (SPV) and Common Spatial Patterns (CSP), and the intermediate EEG image and movie representations~\cite{bashivan2015learning} to this end. We employ classical machine learning and deep learning frameworks such as 1D, 2D and 3D Convolutional Neural Networks (CNN) for emotion decoding. As shown in Fig.~\ref{fig:overview}, we perform (a) categorical and dimensional emotion recognition (binary valence and arousal classification), and (b) PD vs HC recognition from emotional EEG data.  

Key findings from our study are as follows: (1) Dimensional analysis reveals that arousal is better perceived in PD than valence; similar or superior classification is achieved with HC data for both attributes. (2) Fine-grained analyses of emotion class mislabeling reveals confounds among opposite-valence emotions for PD data; this trend is not discernible for HC. (3) Near-ceiling PD vs HC classification (F1 $\geq$ 0.97) is achieved with a 2D-CNN on emotional EEG data, implying that affective neural responses of PD and HC subjects are highly discriminable. Analyzing emotional neural responses can, therefore, enable facile PD diagnosis and treatment. Our study makes the following contributions:     

% The following classification tasks are performed: binary valence classification of PD and HC data as High Valence (HV) and Low Valence (LV){;} binary arousal classification of PD and HC data as High Arousal (HA) and Low Arousal (LA){;} multiclass categorical emotion classification of \emph{{sadness}}, \emph{{happiness}}, \emph{fear}, \emph{disgust}, \emph{surprise} and \emph{anger}{; and} binary classification of the EEG signals of PD patients and HC. 

 %summarises our work, which includes the following contributions:

\begin{itemize}
\item {It is the first {study} to examine (a) PD vs HC recognition, and (b) emotion perception in PD exclusively from EEG classification trends. While resting-state EEG analysis has achieved high PD recognition accuracy~\cite{Shu2020,bhurane2019diagnosis}, it requires EEG acquisition in a highly controlled setting. {In contrast}, we examine EEG signals acquired during the routine task of emotional media consumption, which also allows for PD emotion understanding.}
\item {PD diagnosis and treatment heavily rely on patient self-reports and expert assessments. While the importance of pre-clinical diagnosis and the need for objective monitoring with wearables~\cite{Mantri2019} has been highlighted recently, high PD vs HC discriminability achieved with passive data acquisition during a routine task points to a promising alternative.}
\item {We employ multiple (a) EEG descriptors and (b) machine learning and deep learning recognition frameworks for analyses. Among EEG descriptors, CSPs and SPVs are respectively optimal for emotion and PD recognition. CNN frameworks trained with intermediate EEG image and movie descriptors, however, achieve superior emotion and PD recognition.}

\end{itemize}
% You must have at least 2 lines in the paragraph with the drop letter
% (should never be an issue)
%
%The remainder of the article is organised as follows: Section \ref{related_work} reviews the literature related to emotion perception patterns in PD patients. Section \ref{materials} describes the materials and methods, before Section \ref{val_results}, Section \ref{asl_results}, Section \ref{multi_results}, and Section \ref{pdnc_results}, respectively, detail the results and discussions for binary valence classification, binary arousal classification, multiclass categorical emotion classification and binary PD vs HC classification. Section \ref{conclusion} draws the conclusions.

%
%------------------------------------------------------------
%
\section{Related Work} \label{related_work}
To highlight the novelty of our study, this section reviews related work on (a) emotional impairments in PD patients, and (b) the use of biosignals to assess impairments.

%
%--------------------------------------------------
%
\subsection{Emotional Impairments in PD Patients}
PD patients not only show motor symptoms, but also cognitive~\cite{hsu2012role, hou2007non}, and emotional~\cite{gray2010meta} deficits. Kan~\emph{et al.}~\cite{kan2002recognition} report PD deficits in recognising the fear and disgust facial emotions, while Suzuki~\emph{et al.}~\cite{suzuki2006disgust} observe impaired recognition of disgust. Clark \emph{et al.}~\cite{clark2008specific} note impaired anger recognition in PD patients with left hemisphere pathology, and reduced surprise recognition with right hemisphere pathology. Baggio~\emph{et al.}~\cite{baggio2012structural} observe PD deficits in recognizing sad, anger and disgust, while Narme~\emph{et al.}~\cite{narme2011understanding} note impaired recognition of anger and fear. A meta-analysis indicates an initial PD deficit for negative emotions~\cite{gray2010meta}, and later for positive emotions~\cite{lin2016degraded}.
%the ability of patients to recognise emotions deteriorates from an initial impairment in the negative emotions and then extends the impairment to positive emotions
%use prosodic stimuli, where PD patients had to rate semantically neutral and emotional sentences for emotion.
Some studies employ non-visual stimuli, \eg, auditory and verbal, to assess PD emotion deficits. In an emotional voice  test~\cite{jin2017altered}, PD patients in general exhibit impaired recognition and expression. Kan \emph{et al.}~\cite{kan2002recognition} observe reduced recognition of fear, surprise and disgust from text, and this finding is mirrored by Pell \emph{et al.}~\cite{pell2003processing}. Some studies~\cite{pell2005facial,adolphs1998intact}, however, indicate that PD minimally impacts facial {expression} recognition. %Pell \emph{et al.} and Adolphs \emph{et al.}~\cite{} find limited evidence of impairment as PD patients assess emotional faces.

%
%--------------------------------------------------
%
\subsection{Using Biosignals to Assess Impairments}
Emotion is a psycho-physiological expression related to mood and personality~\cite{DECAF,Ascertain}. Wearable sensing technologies can help examine biosignals and interpret associated emotions. Examining fMRI brain activations  reveals a stronger activation in somatosensory regions  during emotion processing for PD patients~\cite{wabnegger2015facial}. Recognising emotive facial expressions requires somatosensory cortices~\cite{adolphs2000role} connected to the basal ganglia, the primary neurodegeneration site in PD. fMRI analyses show reduced functional activity in the left and right posterior putamen~\cite{moonen2017fmri}, which disturbs emotional processing.

Spontaneous facial expressivity in PD observed via Electromyogram (EMG) and Electrocardiogram (ECG) signals reveal differences between PD patients and controls~\cite{wu2014objectifying}. Examining eye movements, PD patients are found to make fewer fixations while viewing affective scenes~\cite{dietz2011emotion}. Dietz~\emph{et al.} found ocular movements to be more compromised than pupil dilation in PD due to the disruption in basal ganglia circuitry. Measuring eye blinks via EMG, an attenuated reactivity to aversive stimuli is observed in PD due to an amygdala-based translational defect~\cite{bowers2006startling}. A neuroimaging study involving fearful faces notes that dopamine levels modulate {the} amygdala's response in PD~\cite{tessitore2002dopamine}, and amygdala dysfunction induces impaired reactions to fear-inducing stimuli. Miller \emph{et al.}~\cite{miller2009startle} report that PD patients show reduced reactivity to highly arousing negative stimuli. 

%
%--------------------------------------------------
%
%\subsection{Physiological Feature Extraction and Classification}
Deep learning has become popular for a variety of pattern recognition tasks including  neural applications~\cite{krizhevsky2012imagenet, graves2013speech, karpathy2014large}. CNNs have been employed for EEG-based seizure prediction~\cite{mirowski2009classification}, Alzheimer's detection~\cite{besthorn1997discrimination}, fMRI-based schizophrenia detection~\cite{qureshi20193d}, \etc. Deep learning methods learn salient and latent neural representations~\cite{plis2014deep}. EEG-based categorical emotion recognition in PD patients has been pursued via higher-order spectral statistics~\cite{yuvaraj2014emotion,yuvaraj2014optimal,yuvaraj2016brain,yuvaraj2016hemispheric}. PD recognition via spectral analysis of resting-state EEG has been performed with \emph{k}-Nearest Neighbour and Support Vector Machine classifiers~\cite{bhurane2019diagnosis}. A 13-layer 1D-CNN for PD vs HC classification with resting-state EEG is proposed in \cite{Shu2020}. Binary  PD vs HC EEG classification using a convolutional-recurrent neural network is proposed in~\cite{lee2019deep}.%

%
%--------------------------------------------------
%
\begin{table*}[!ht]
\centering
\fontsize{7}{7}\selectfont
\renewcommand{\arraystretch}{1.2}
\caption{Overview of studies examining PD behavior and automated PD detection.}\vspace{-1mm}
\label{rw_comp}
%\scriptsize
    \begin{tabular}{|c|c|p{4.5cm}|p{4.5cm}|p{3.5cm}|}
     \hline
     \multirow{2}{*}{\centering{\textbf{Behavior Studied}}}&
     \multirow{2}{*}{\textbf{\centering{References}}}&
     \multirow{2}{*}{\textbf{\hspace{1.5cm}{Description}}} & \multirow{2}{*}{\textbf{\hspace{1.5cm}{Findings}}} & 
     \multirow{2}{*}{\textbf{\hspace{1.5cm}{Remarks}}}\\ 
     &  & & & \\ \hline
     \multirow{4}{*}{Explicit+Implicit} &  \multirow{4}{*}{\cite{smith1996spontaneous}} & PD and normal subjects shown emotional video clips and their emotional reactions encoded via Facial Action Coding System. Emotional self-ratings compiled. & PD group showed significantly less facial activity than controls. & Purely behavioral study. Automated PD detection not attempted.\\ \hline
     \multirow{3}{*}{Implicit} & \multirow{3}{*}{\cite{simons2003disturbance}} & PD and normal subjects presented with odours and their emotional reactions encoded via Facial Action Coding System. & Spontaneous facial activity disturbed in PD, and also impaired ability to pose and mask facial expressions. & Purely behavioral study. Automated PD detection not attempted. \\ \hline
     \multirow{3}{*}{Implicit} & \multirow{3}{*}{\cite{jacobs1995emotional}} & PD and controls presented with emotional faces, and were asked to pose those expressions. & PD patients were impaired relative to controls on making emotional faces. & Purely behavioral study. Automated PD detection not attempted. \\ \hline
     \multirow{4}{*}{Explicit} & \multirow{4}{*}{\cite{paulmann2010dynamic}} & Emotion recognition for PD and controls compared via standardized tests across three information channels: lexical-semantic, prosody, and facial. & PD emotion recognition capability increased with more channels, but PD group performed worse than controls across all channels. & Behavioral study with no automated PD detection. \\ \hline
     \multirow{4}{*}{Explicit+Implicit} & \multirow{4}{*}{\cite{miller2009startle}} & 24 PD and 24 HC subjects presented with emotional pictures, and responses acquired via self-ratings (explicit) and EMG activity (implicit). & Reduced PD reactivity to low-valence, high-arousal pictures; behavior was not specific to any emotion category (\eg, fear, disgust). & Behavioral study with no automated PD detection.  \\
     \hline
     \multirow{4}{*}{Implicit (Rest-state EEG)} &  \multirow{4}{*}{\cite{bhurane2019diagnosis,Shu2020}} &  PD vs HC recognition via resting-state EEG analysis using machine learning~\cite{bhurane2019diagnosis} and deep learning~\cite{Shu2020} approaches. Dataset proposed in~\cite{bhurane2019diagnosis} examined.  & Mean recognition accuracy of 99.1\% achieved with SVM~\cite{bhurane2019diagnosis}, and 88.3\% achieved with 1D-CNN~\cite{Shu2020}. & Resting-state EEG compiled under highly controlled conditions, ecologically invalid.\\ \hline
\multirow{3}{*}{Implicit (Emotional EEG)} & \multirow{3}{*}{\cite{YUVARAJ2014108}} & PD vs HC recognition via emotional EEG analysis with SVM classification. Dataset proposed in~\cite{yuvaraj2014emotion} examined. & Mean accuracy of 87.9\% achieved. & Routine and ecologically valid setting. Only PD vs HC classification attempted. \\ \hline
\multirow{5}{*}{Implicit (Emotional EEG)} & \multirow{5}{*}{\textbf{Our work}} & \vspace{1mm} Dimensional and categorical emotion plus PD recognition via machine and deep learning approaches. Dataset in ~\cite{yuvaraj2014emotion} examined. & Valence-specific mislabeling observed with PD data, while no arousal-related differences noted between PD and HC groups. Maximum accuracy/F1-score of 93\%, 98\% and 99\% achieved for valence, arousal and PD recognition, respectively. & \textbf{PD and emotion EEG recognition with multiple classification methods. PD recognition higher accuracy with emotional EEG, and comparable to prior work with resting-state EEG.} \\ \hline
    \end{tabular}
		\vspace{-3mm}
\end{table*}

\subsection{Identifying Research Gaps}
While it is largely known that PD patients face emotional deficits, only a few studies examine these deficits via implicitly acquired biosignals such as EEG. Moreover, prior studies on automated PD diagnosis have only examined resting-state EEG, but not emotional EEG signals. Emotional EEG signals can be (a) acquired via easy-to-use, portable headsets under routine settings, and (b) utilized for PD recognition as well as studying PD emotion deficits as described in this work. Table~\ref{rw_comp} compares and contrasts our work against the literature; evidently, our approach allows for state-of-the-art PD recognition with ecologically valid data, different from prior work on resting-state and emotional EEG. 

%To address these limitations, we examine emotion perception in PD patients by employing both machine learning and deep learning algorithms for dimensional and categorical emotion classification from PD and HC data. These analyses allow for a fine-grained investigation of how PD impacts emotion perception. We also perform binary PD vs HC classification from emotional state EEG signals to show that emotional EEG responses encode PD characteristics. 

%
%--------------------------------------------------
%
\section{Materials and Methods}\label{materials}
\begin{figure*}[t]
    \centering
    \includegraphics[width=0.9\textwidth]{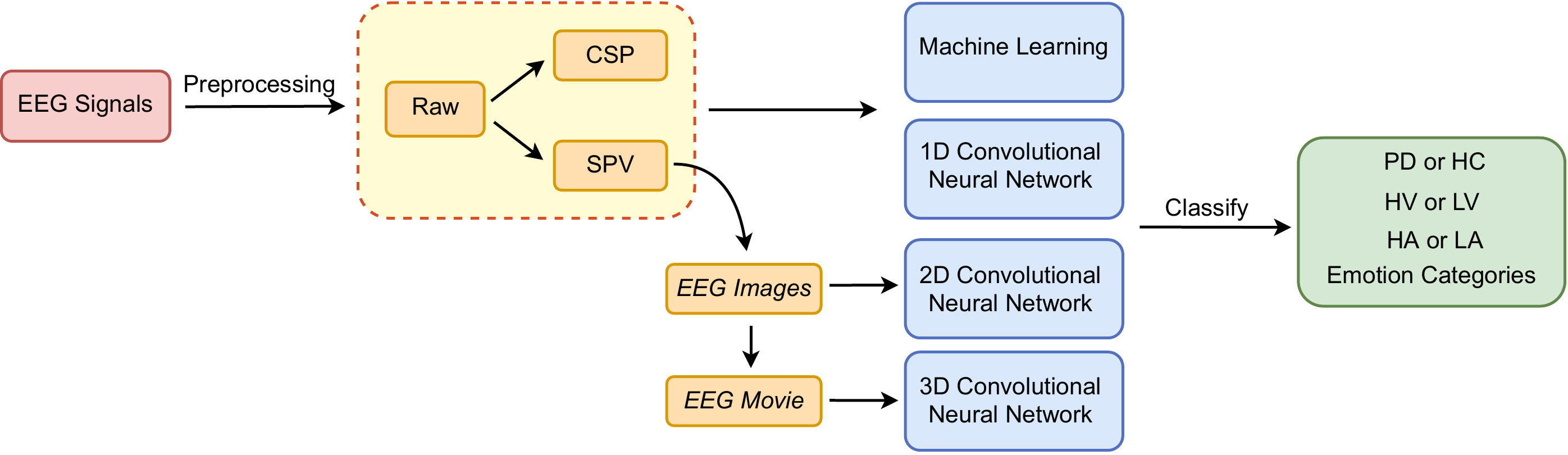}\vspace{-2mm}
    \caption{\textbf{Overview:} Our pipeline involves (a) EEG pre-processing and extraction of low-level features such as spectral power vectors (SPV) and common spatial patterns (CSP), (b) feeding of these features or intermediate representations such as EEG images and movies to machine and deep learning frameworks to perform (i) dimensional and discrete emotion recognition, and (ii) PD vs HC classification. 
    } 
    \label{fig:overview}\vspace{-4mm}
\end{figure*}
\subsection{Dataset}\label{Dataset}
The dataset comprises EEG signals from 20 non-demented PD (10 males/10 females) and 20 HC (9 males/11 females) subjects from Hospital Universiti Kebangsaan Malaysia, Kuala Lumpur, upon ethics approval (approval no. UKM1.5.3.5/244/FF-354-2012)~\cite{yuvaraj2014emotion,yuvaraj2016hemispheric}. EEG data was recorded via the 14-channel wireless \emph{Emotiv Epoc} headset (128 Hz sampling rate). Audio-visual stimuli are used to induce the six Ekman emotions (\emph{sadness}, \emph{happiness}, \emph{fear}, \emph{disgust}, \emph{surprise}, \emph{anger}), resulting in a total of 1440 samples (2 classes $\times$ 20 subjects $\times$ 6 emotions $\times$ 6 trials/emotion).
Stimuli from the IADS~\cite{lang2007} (audio) and IAPS~\cite{lang1997} (visual) datasets were combined. Each trial (stimulus viewing episode) lasts 4-5 min. PD patients were optimally medicated to reduce tremors, and informed consent obtained from all participants. Data acquisition, PD clinical history, ethics approval, PD inclusion and exclusion criteria are described in~\cite{yuvaraj2014emotion,yuvaraj2016brain}. 

%
%----------------------------------------
%
\subsection{Data Preprocessing} \label{Data Preprocessing}
EEG outlier samples were discarded by limiting the signal amplitude to $85 \mu V$, followed by an IIR bandpass Butterworth filter to retain the $8-49$~Hz range~\cite{yuvaraj2016brain}. The filtered EEG signal, termed as \emph{raw} data henceforth, was segmented into 5-second epochs to preserve temporal information following~\cite{kim2007bimodal}. Thus, $1440$ original EEG samples were segmented into 13193 epochs for feature extraction. For classical machine learning methods, raw features were $z$-normalised followed by Principal Component Analysis (PCA) to retain $95\%$ data variance (PCA was not part of the CNN pipeline). 

% The study conducted in  proposed that the window size depends on the modality, i.e., 2–6 s for speech, and 3–15 s for biosignals. Based on this, we segment the EEG signals into  and extract features from each epoch separately ensuring a conservation of , rather than averaging the epochs into a single slice of activity map. By slicing the 1440 EEG signals of roughly 50-55 seconds, we obtain a total of  samples.

%
%----------------------------------------
%
\subsection{Feature Extraction from raw EEG} \label{Feature Extraction}
%The below features were extracted from the raw EEG signal.

%
%------------------------------
%
\subsubsection{Spectral Power Vector (SPV)}
Power Spectral Analysis was performed to estimate EEG spectral density upon spectral transformation~\cite{roy2019deep}. On each epoch, a Butterworth bandpass filter was applied to extract the $\alpha$ ($8-13$ Hz), $\beta$ ($13-30$ Hz) and $\gamma$ ($30-49$ Hz) spectral bands. A Fast Fourier Transform (FFT) was performed, followed by summation of squared FFT values within each three frequency band over the 14 electrodes to obtain the concatenated spectral power vector $[\alpha_1, \beta_1$, $\gamma_1$, \ldots, $\alpha_{14}, \beta_{14}, \gamma_{14}]$. %These values were concatenated to obtain the  \alpha_1, \alpha_2, \ldots, \alpha_{14}, \beta_1, \beta_2, \ldots, \beta_{14}, \gamma_1, \gamma_2, \ldots, \gamma_{14}]$.  

%
%------------------------------
%
\subsubsection{Common Spatial Patterns (CSP)}
Common Spatial Patterns were extracted by learning a linear combination of the original features~\cite{ramoser2000optimal}. Filters (transformations) were designed so that the transformed signal variance was maximal for one class and minimal for the other. Apart from dimensionality reduction, CSPs enable recovery of the original signal by gathering relevant information spread over multiple channels and are, hence, popular EEG features~\cite{roy2019deep}. We learn the spatial transform $w$, which maximises the function:
\begin{equation}
    J_{CSP}(w) = \dfrac{wX_{1}X_{1}^{T}w^T}{wX_{2}X_{2}^{T}w^T} = \dfrac{wC_{1}w^T}{wC_{2}w^T}
    \label{eq::J_CSP}
\end{equation}
where $C_i, X_i$ are, respectively, the spatial covariance matrix and the bandpass-filtered signal matrix for class $i$. In Eq.~\ref{eq::J_CSP}, $wX_{i}$ is the spatially filtered EEG signal for class $i$ and $wX_{i}X_{i}^{T}w^T$ is the the transformed signal variance, \ie, band power of the filtered signal. Thus, maximising $J_{CSP}(w)$ leads to spatially filtered signals whose inter-class band power ratio is maximum, and can be solved via eigenvalue decomposition. The spatial filters $w$ that maximise $J_{CSP}(w)$ are the eigenvectors of the highest and lowest eigenvalues for matrices $C_1, C_2$. Hence, $w$ gives feature vectors that are optimal for  class discrimination. For each class $i$, the variances of only a small number of signals most suitable for distinguishing are used. Six filters corresponding to the three largest and smallest eigenvalues are used for generating the CSP feature $$ f = \log(wC_{2}w^T) = \log(var(wX_2))$$
of dimension $(1,6)$ for each $(14,640)$ EEG epoch. It is shown that the classification accuracy does not improve with larger number of filters \cite{muller1999designing}.

%. Hence, each $5$s EEG epoch of dimension  generated a . 

%This process is repeated for all the epochs in the sample and subsequently for all the EEG samples.

%Once these filters are obtained, a CSP feature $f$ is defined as $ f = \log(wC_{2}w^T) = \log(var(wX))$. 
%

%   $$
%
%
%----------------------------------------
%
\subsection{Classical Machine Learning (ML) Algorithms}
Raw EEG data or extracted features were input to machine/deep learning classifiers (Fig.~\ref{fig:overview}). We explored the following ML algorithms. 
\begin{itemize}
    \item $k$-Nearest Neighbour (kNN), where the test sample is assigned the label corresponding to the mode of its $k$-\textit{closest} neighbours based on a suitable distance metric.
    \item Support Vector Machine (SVM), where input data are transformed to a high-dimensional space where the two classes are linearly separable and the inter-class distance is maximum.   
    \item Gaussian Naive Bayes (GNB), a generative classifier assuming class-conditional feature independence. 
    \item Decision Tree (DT), which uses a tree-like graph structure where each leaf node represents a category label. 
    \item Linear Discriminant Analysis (LDA), which linearly transforms data to achieve maximal inter-class distance. 
    \item Logistic Regression (LR), which maps the input to class labels via the sigmoid function.   
\end{itemize}

%
%----------------------------------------
%
\subsection{Convolutional Neural Network Pipeline} \label{deep}
Deep neural networks are the state-of-the-art in text, speech, image, video and EEG-based recognition~\cite{graves2013speech, hermann2015teaching, karpathy2014large, krizhevsky2012imagenet,bashivan2015learning,Shukla2020}, and
have outperformed traditional machine learning methods obviating the need for handcrafted features~\cite{roy2019deep}. 
We explored 1D, 2D and 3D-CNNs to learn EEG representations. Raw or extracted EEG features were fed to the 1D-CNN; feature dimensions input to the 1D-CNN with raw, spectral and CSP descriptors were, respectively, (640, 14), 42 and 6. However, this representation ignores the EEG spatial structure; therefore, we synthesised the \emph{EEG image} and \emph{EEG movie} descriptors to preserve the spatial structure.

Extracted SPVs were transformed to an EEG image as in~\cite{bashivan2015learning}. EEG electrodes, distributed on the scalp in 3D were projected onto a 2D surface to capture the spatial activity distribution. Azimuthal Equidistant Projection was used to preserve the relative inter-electrode distance~\cite{snyder1987map}. Scattered scalp power measurements were interpolated to derive a $32 \times 32$ pixel EEG image. Repeating this process for the $\alpha, \beta$ and $\gamma$ bands produced three topo-maps, which were then merged to form a 3-channel ($32 \times 32 \times 3$) EEG image. To learn the temporal EEG structure, given that 3D-CNNs effectively learn from video chunks~\cite{yue2015beyond,Chugh20}, we synthesised EEG movie samples comprising five images generated by sliding non-overlapping $1s$ windows over the $5s$ epoch. The 3D-CNN input dimensionality is $5 \times 32 \times 32 \times 3$.

% CNNs are established to be powerful models to classify not only single static images, but a series of images given as frames. 3D-CNNs are employed for large-scale video classification tasks, where the model is exposed to not just the spatial information, but the temporal structure as well . In this study, the input data for 3D-CNNs are a sequence of images stacked adjacent to each other. In a non-overlapping 5-second slice of the preprocessed EEG sample, The

\begin{figure}[!htbp]
    \centering
    \includegraphics[width=0.45\textwidth]{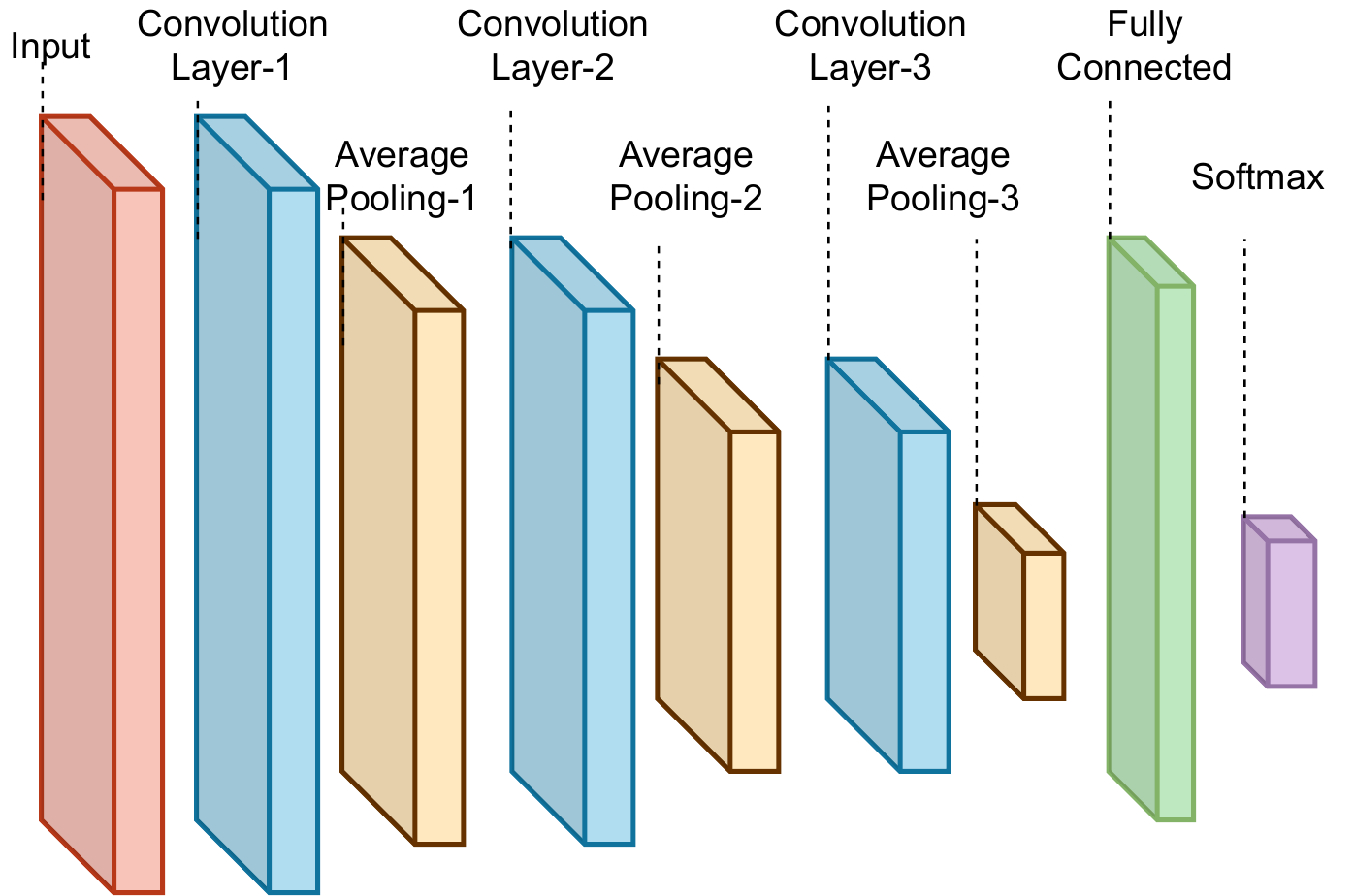}\vspace{-2mm}
    \caption{Basic architecture of the 1D/2D/3D-CNN.}\vspace{-3mm}
    \label{fig:cnn_arch}
\end{figure}

The general architecture of the three-layered 1D/2D/3D-CNN employed for classification is shown in Fig.~\ref{fig:cnn_arch}. Output dimensions for each CNN layer are presented in Table~\ref{tab:arch_dimensions}. Three convolutional layers convolve the input signal with a stride of 3, and comprise 16, 32 and 32 filters of size 3, $3\times3$  and $3\times3\times3$, respectively. Each convolutional layer is followed by average pooling over 2-unit regions. Batch normalisation is applied to normalise prior activations, and a dropout of $0.1-0.5$ is employed for regularisation. The dense layer comprises 128 neurons, followed by a softmax layer composed of {two} neurons (for dimensional emotion and PD vs HC classification) or 6 neurons (for categorical emotion recognition) conveying class 
probabilities. CNN hyper-parameters (learning rate $\in [10^{-5}, \ldots,10^{-1}$], optimizer $\in \{\text{SGD, Adam, RMS Propogation}\}$ and dropout rate) were tuned via 10-fold cross-validation identical to the ML models\footnote{Code available at \url{https://github.com/ravikiranrao/PD-EEG}}. 

\subsection{Performance Evaluation:} All models were fine-tuned via exhaustive grid-search and performance evaluated via ten-fold cross-validation. For all results, we report the \textbf{weighted F1}  measure or the weighted mean of the per-class F1-scores, which accounts for class imbalance noted in valence and arousal classification.
%

% Please add the following required packages to your document preamble:
% \usepackage{multirow}
% \usepackage[table,xcdraw]{xcolor}
% If you use beamer only pass "xcolor=table" option, i.e. \documentclass[xcolor=table]{beamer}
\begin{table}[!ht]
\caption{Output dimensions of each layer in 1D, 2D and 3D-CNN.}\vspace{-2mm}
\resizebox{0.5\textwidth}{!}{%
\begin{tabular}{c|ccc|c|c}
                        & \multicolumn{3}{c|}{\textbf{1D CNN}} &                          &                          \\ \cline{2-4}
\multirow{-2}{*}{\textbf{Layer}} & \textbf{Spectral} & \textbf{CSP} &  \textbf{Raw}     & \multirow{-2}{*}{\textbf{2D CNN}} & \multirow{-2}{*}{\textbf{3D CNN}} \\ \hline
\rowcolor[HTML]{B1DDF0} 
Convolution layer - 1   & 42, 16   & 6, 16  & 640, 16 & 32, 32, 16               & 5, 32, 32, 16            \\
\rowcolor[HTML]{FFE8BF} 
Average Pooling - 1     & 21, 16   & 3, 16  & 320, 16 & 16, 16, 16               & 3, 16, 16, 16            \\
\rowcolor[HTML]{B1DDF0} 
Convolution layer - 2   & 21, 32   & 3, 32  & 320, 32 & 16, 16, 32               & 3, 16, 16, 32            \\
\rowcolor[HTML]{FFE8BF} 
Average Pooling - 2     & 11, 32   & 2, 32  & 160, 32 & 8, 8, 32                 & 2, 8, 8, 32              \\
\rowcolor[HTML]{B1DDF0} 
Convolution layer - 3   & 11, 32   & 2, 32  & 160, 32 & 8, 8, 32                 & 2, 8, 8, 32              \\
\rowcolor[HTML]{FFE8BF} 
Average Pooling - 3     & 6, 32    & 1, 32  & 80, 32  & 4, 4, 32                 & 1, 4, 4, 32              \\
Flatten                 & 192      & 32     & 2560    & 512                      & 512                      \\
Batch Normalisation     & 192      & 32     & 2560    & 512                      & 512                      \\
\rowcolor[HTML]{BEE8C0} 
Fully Connected         & 128      & 128    & 128     & 128                      & 128                      \\
\rowcolor[HTML]{DDC2E7} 
Softmax                 & 2 or 6   & 2 or 6 & 2 or 6  & 2 or 6                   & 2 or 6                   \\ \hline
\end{tabular}%
}
\hspace{5mm}
\label{tab:arch_dimensions}\vspace{-4mm}
\end{table}

% Please add the following required packages to your document preamble:
% \usepackage{multirow}
\begin{table*}[!ht]
\centering
\fontsize{7}{7}\selectfont
\caption{\textbf{Binary valence classification:} F1-scores are of the form of $\mu \pm \sigma$. ML results denote the best among kNN, SVM, GNB, LDA and LR.} \vspace{-2mm}
\resizebox{2\columnwidth}{!}{%
\begin{tabular}{c|ccc|ccc|c|c}
%\hline
\multirow{2}{*}{\textbf{Data}} & \multicolumn{3}{c|}{\textbf{ML}}                    & \multicolumn{3}{c|}{\textbf{1D-CNN}}                & \multirow{2}{*}{\textbf{2D-CNN}} & \multirow{2}{*}{\textbf{3D-CNN}} \\ \cline{2-7}
                               & \textbf{SPV} & \textbf{CSP}   & \textbf{Raw}   & \textbf{SPV} & \textbf{CSP}   & \textbf{Raw}   &                                  &                                  \\ \hline
Full                           & 0.78 $\pm$ 0.01    & 0.81 $\pm$ 0.01 & 0.55 $\pm$ 0.01 & 0.85 $\pm$ 0.01    & 0.82 $\pm$ 0.02 & 0.88 $\pm$ 0.11 & 0.89 $\pm$ 0.06                   & 0.91 $\pm$ 0.05                   \\
PD                             & 0.75 $\pm$ 0.02     & 0.84 $\pm$ 0.01  & 0.55 $\pm$ 0.01 & 0.82 $\pm$ 0.09    & 0.86 $\pm$ 0.03 & 0.75 $\pm$ 0.17 & 0.86 $\pm$ 0.07                   & 0.91 $\pm$ 0.07                   \\
HC                             & 0.81 $\pm$ 0.02    & 0.84 $\pm$ 0.01  & 0.56 $\pm$ 0.01  & 0.85 $\pm$ 0.08    & 0.88 $\pm$ 0.04 & 0.90 $\pm$ 0.11 & 0.91 $\pm$ 0.07                   & \textbf{0.93 $\pm$ 0.05}                   \\ \hline
\end{tabular}%
}
\hspace{5mm}
\label{tab:HVLV_Table}\vspace{-2mm}
\end{table*}

\begin{figure*}[!ht]
%\captionsetup[subfigure]{aboveskip=-1pt,belowskip=-1pt}
\centering
\begin{subfigure}
  \centering
  \includegraphics[width=.32\linewidth]{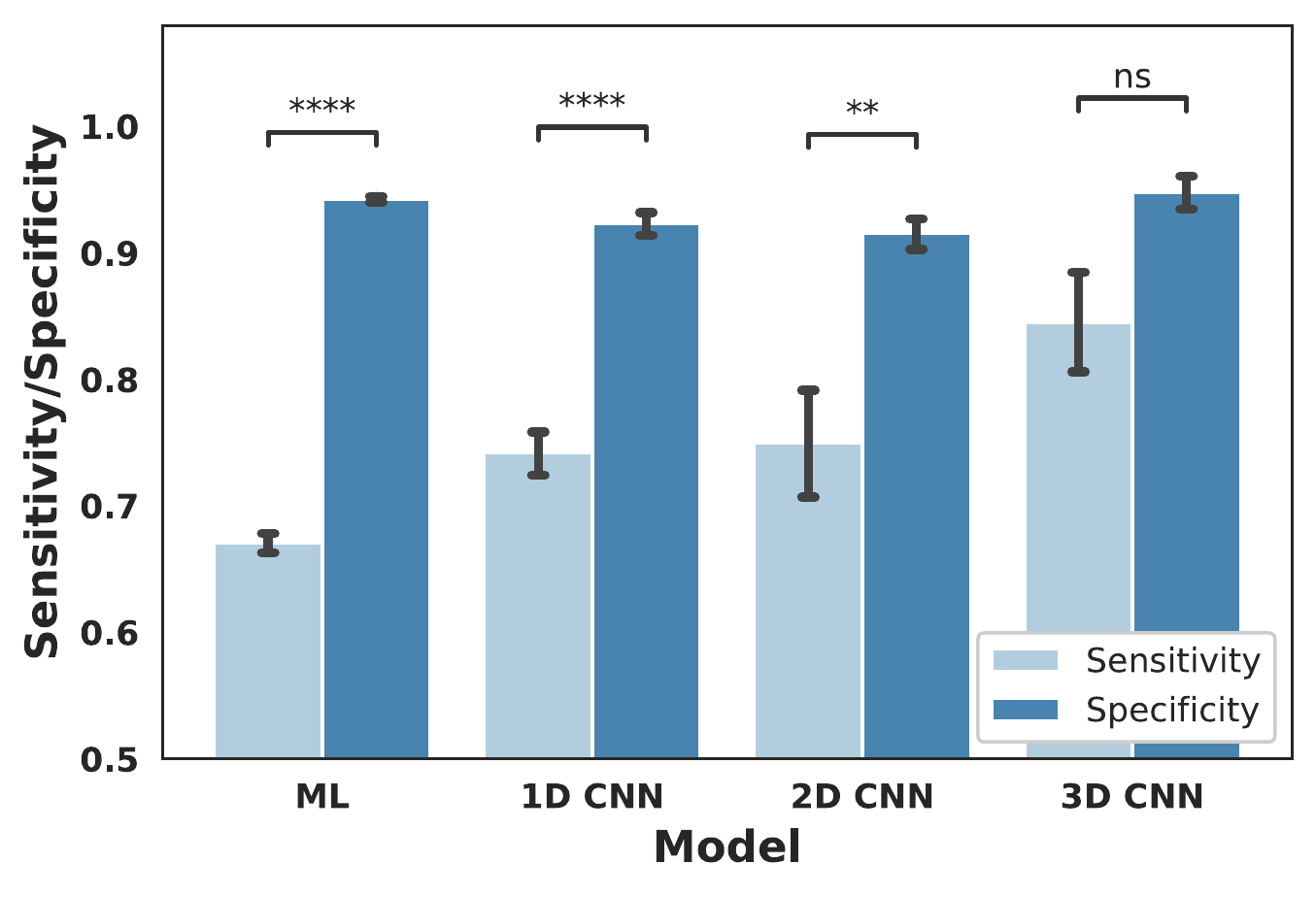}
  %\caption{A subfigure}
  %\label{fig:sub1}
\end{subfigure}%
%\hspace{-2mm}
\begin{subfigure}
  \centering
  \includegraphics[width=.32\linewidth]{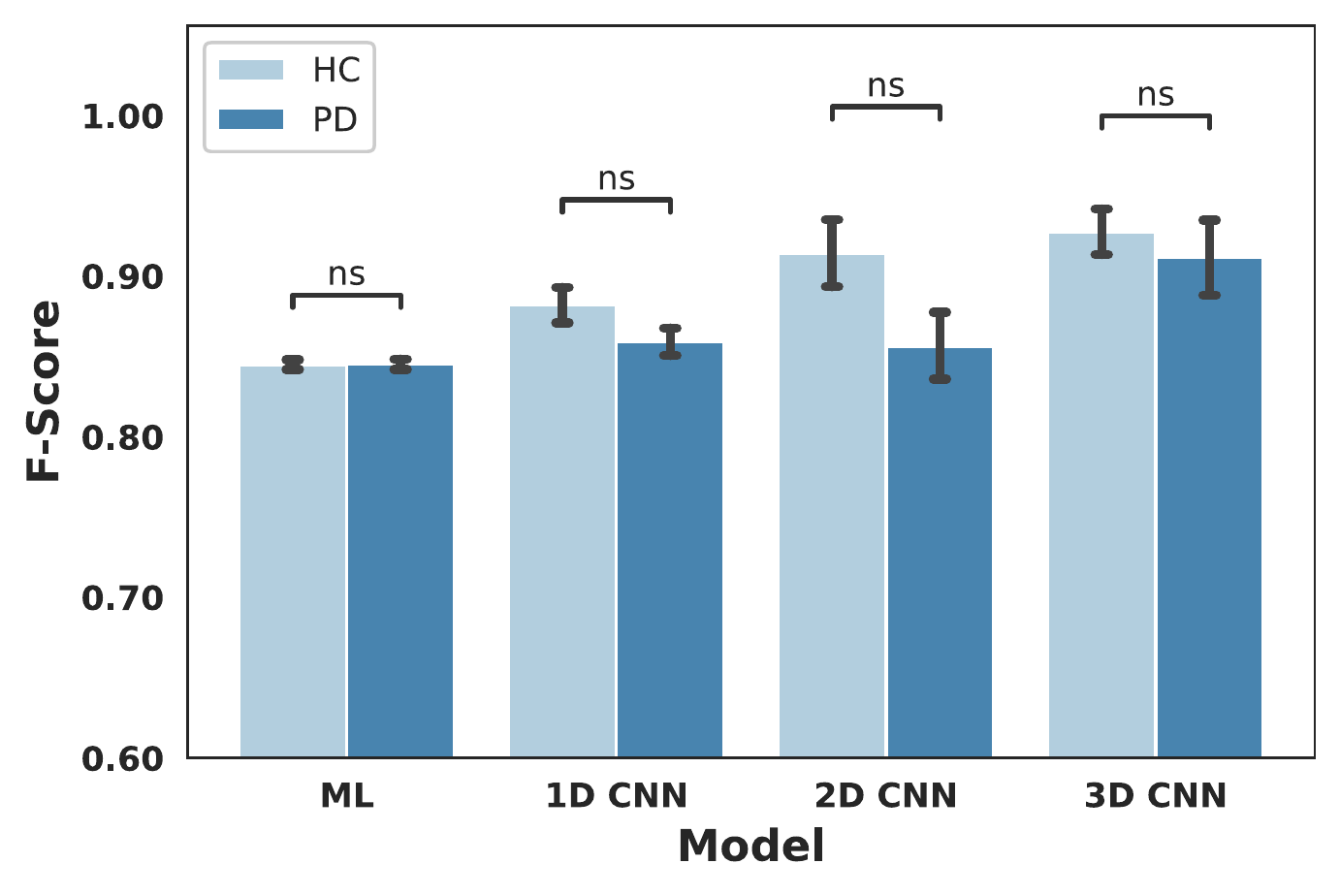}
  %\caption{A subfigure}
  %\label{fig:sub2}
\end{subfigure}
\begin{subfigure}
  \centering
  \includegraphics[width=.32\linewidth]{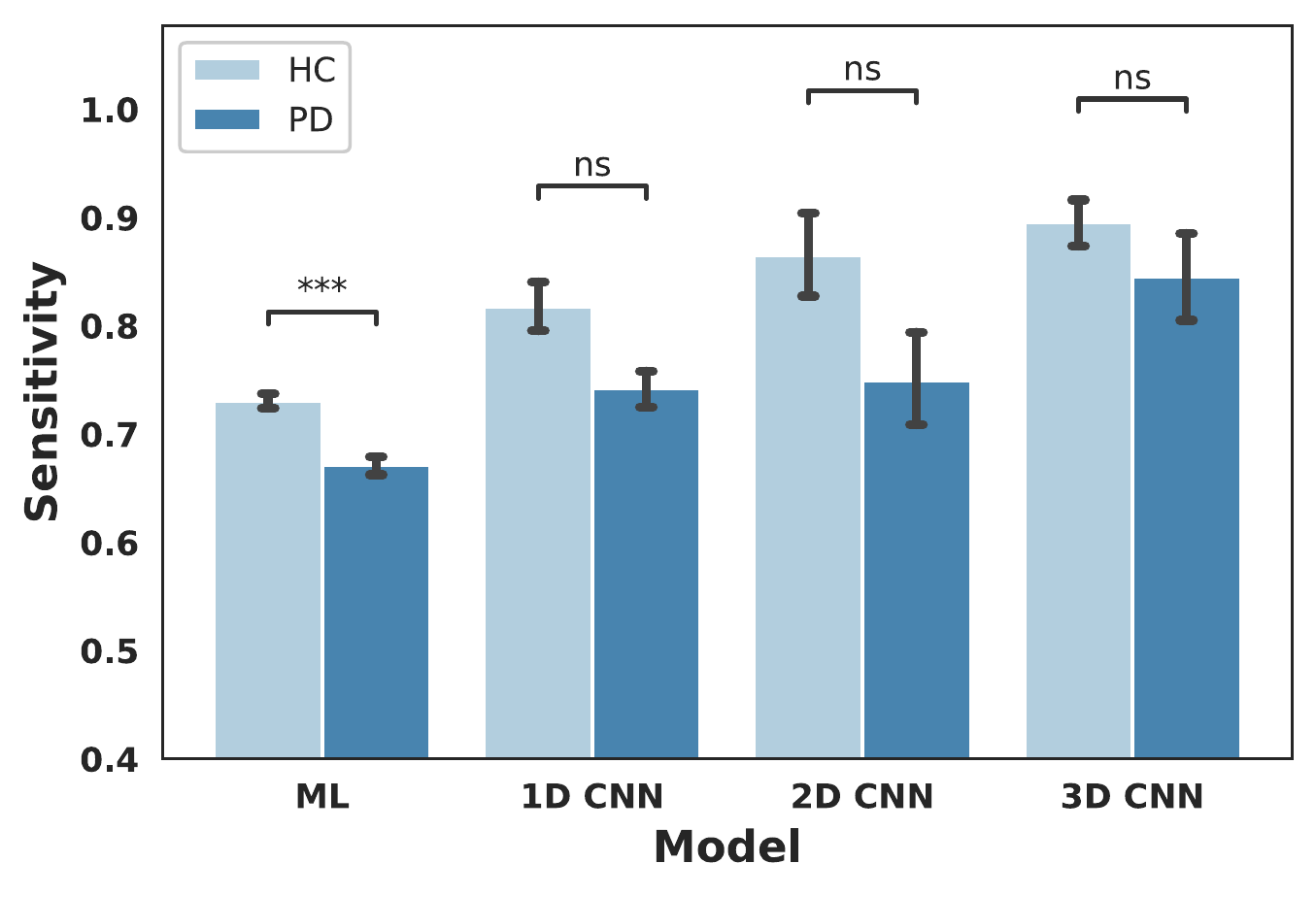}
  %\caption{A subfigure}
  %\label{fig:sub2}
\end{subfigure}
\vspace{-2mm}
\caption{\textbf{Binary valence classification:} (Left) Sensitivity and specificity with PD data across various models. (Centre) F1-scores with PD and HC data across various models. (Right) Sensitivity on PD and HC data across various models. Error bars denote standard error of mean (SEM). **** $\implies p < 0.0001$, *** $\implies p < 0.001$, ** $\implies p < 0.01$, * $\implies p < 0.05$ and n.s.\ $\implies p > 0.05$ as per a Tukey HSD test.}
\label{fig:HVLV_fig}\vspace{-4mm}
\end{figure*}

% While the input data for traditional ML algorithms and 1D-CNN are the hand-crafted features, 2D-CNN captures the spatial information inherent in the data, which was ignored when the data was fed as feature vectors. 
%In Sections \ref{val_results} to \ref{pdnc_results}, we present the results and discussions corresponding to the various classification tasks.

%
%--------------------------------------------------
%
\section{Experiments \& Results}\label{exp_results}
\subsection{Valence Classification}\label{val_results}
To examine emotional perception in PD, we first performed valence classification by training binary classifiers with (a) PD, (b) HC, and (c) PD$+$HC or \textit{full} EEG data. \emph{Happiness} and \emph{surprise} were grouped in the high valence (HV) category, while \emph{sadness}, \emph{fear}, \emph{disgust} and \emph{anger} data were grouped in the low valence (LV) category. The HV:LV class ratio within PD, HC and full data is 1:2.

%
%----------------------------------------
%
\subsubsection{Results} \label{results_hvlv}
Table \ref{tab:HVLV_Table} presents valence classification results on full, PD and HC data with various models. Higher F1-scores were achieved with HC data, implying reduced discriminability with PD EEG data. We discuss the results below.

%
%----------------------------------------
%
\paragraph{Classification with PD Data}
The impact of descriptors on the efficacy of ML techniques is evident from Table~\ref{tab:HVLV_Table}. A one-way Analysis of Variance (ANOVA) to examine the effect of features (Raw, SPV and CSP) on F1-scores revealed confirmed the impact of descriptor type $(F(2,27) = 969.05, p < 0.0001)$. Comparing F1-scores from ten classifier runs, post-hoc Tukey tests revealed significant differences between predictive powers of SPV vs CSP $(p<0.001)$, CSP vs Raw $(p<0.001)$, and SPV vs Raw features $(p<0.001)$. Maximum F1-scores were achieved with CSPs. 

Higher F1-scores were observed with the 1D-CNN for all features, revealing the superior learning ability of CNNs. A one-way ANOVA revealed the minimal impact of different features on 1D-CNN performance, even as CSP features achieved the highest F1 of 0.86. 2D and 3D-CNN achieved even higher F1-scores, conveying that the EEG-image and movie representations are most efficient for valence prediction. The 3D-CNN achieved the maximum F1-score of 0.91.

Fig.~\ref{fig:HVLV_fig} (left) presents model sensitivity and specificity with PD data. \emph{Sensitivity} denotes the true positive rate or proportion of HV samples classified correctly, while \emph{Specificity} denotes the true negative rate or the proportion of correctly classified LV samples. With ML algorithms, a significantly higher mean specificity (0.94) was observed than sensitivity (0.67). A similar trend was observed with 1D and 2D-CNN, with much higher specificity scores noted in both cases. Comparable mean specificity (0.95) and sensitivity (0.85) scores were noted, however, with the 3D-CNN. Overall, these trends convey reduced positive valence recognition with PD data.

%
%----------------------------------------
%
\paragraph{Classification using HC Data}
Higher F1-scores were obtained on HC data for all features and methods (Table~\ref{tab:HVLV_Table}). The impact of features on ML performance was confirmed by an ANOVA test $(F(2,27) = 1382.90, p < 0.0001)$, with CSP features outperforming SPV and Raw features. Higher F1-scores were noted with the 1D-CNN, with all features performing similarly. 2D and 3D-CNN performed better than the 1D-CNN, with the 3D-CNN achieving the best mean F1-score of 0.93.

%
%----------------------------------------
%
\paragraph{PD vs HC F1 Comparison}
Fig.~\ref{fig:HVLV_fig} (centre) compares the valence F1-scores obtained with PD and HC data over all models, with CSP scores plotted for the ML and 1D-CNN methods. While identical scores were achieved on PD and HC data employing ML methods, marginally higher F1-scores were noted on HC data with the 1D, 2D and 3D-CNN. Overall, better classification was achieved with HC rather than PD data. Fig.~\ref{fig:HVLV_fig} (right) compares sensitivity with PD data and HC data across models. With ML algorithms, sensitivity on PD data ($0.67$) was significantly lower than HC ($0.73$) as per a $t$-test $(t(18) = 5.39, p < 0.0001)$. Lower sensitivity scores were again noted on PD data with the 1D, 2D and 3D-CNN even if the differences were insignificant. Cumulatively, these results reveal lower sensitivity for PD EEG data. 

\begin{table*}[!t]
\centering
\fontsize{7}{7}\selectfont
\caption{\textbf{Binary arousal classification:} F1-scores are of the form $\mu \pm \sigma$. Best ML results were achieved with the kNN or NB algorithms. NB results are denoted using a * symbol.}\vspace{-2mm}
\resizebox{2\columnwidth}{!}{%
\begin{tabular}{c|ccc|ccc|c|c}
%\hline
\multirow{2}{*}{\textbf{Data}} &
  \multicolumn{3}{c|}{\textbf{ML}} &
  \multicolumn{3}{c|}{\textbf{1D CNN}} &
  \multirow{2}{*}{\textbf{2D CNN}} &
  \multirow{2}{*}{\textbf{3D CNN}} \\ \cline{2-7}
     & \textbf{SPV} & \textbf{CSP}   & \textbf{Raw}     & \textbf{SPV} & \textbf{CSP}     & \textbf{Raw}     &                  &                  \\ \hline
Full & 0.92 $\pm$ 0.01    & 0.93 $\pm$ 0.00  & 0.76 $\pm$ 0.00* & 0.95 $\pm$ 0.03  & 0.92 $\pm$ 0.01 & 0.95 $\pm$ 0.07 & 0.97 $\pm$ 0.02 & 0.97 $\pm$ 0.02 \\
PD   & 0.92 $\pm$ 0.01    & 0.93 $\pm$ 0.01 & 0.76 $\pm$ 0.00 & 0.96 $\pm$ 0.03  & 0.92 $\pm$ 0.02 & 0.94 $\pm$ 0.06 & \textbf{0.98 $\pm$ 0.02} & 0.98 $\pm$ 0.03 \\
HC   & 0.91 $\pm$ 0.01    & 0.94 $\pm$ 0.01 & 0.76 $\pm$ 0.00* & 0.91 $\pm$ 0.05  & 0.95 $\pm$ 0.01 & 0.94 $\pm$ 0.07 & 0.94 $\pm$ 0.03 & 0.97 $\pm$ 0.02 \\ \hline
\end{tabular}%
}
\hspace{5mm}
\label{tab:HALA_Table} \vspace{-2mm}
\end{table*}

\begin{figure*}[!t]
%\captionsetup[subfigure]{aboveskip=-1pt,belowskip=-1pt}
\centering
\begin{subfigure}
  \centering
  \includegraphics[width=.32\linewidth]{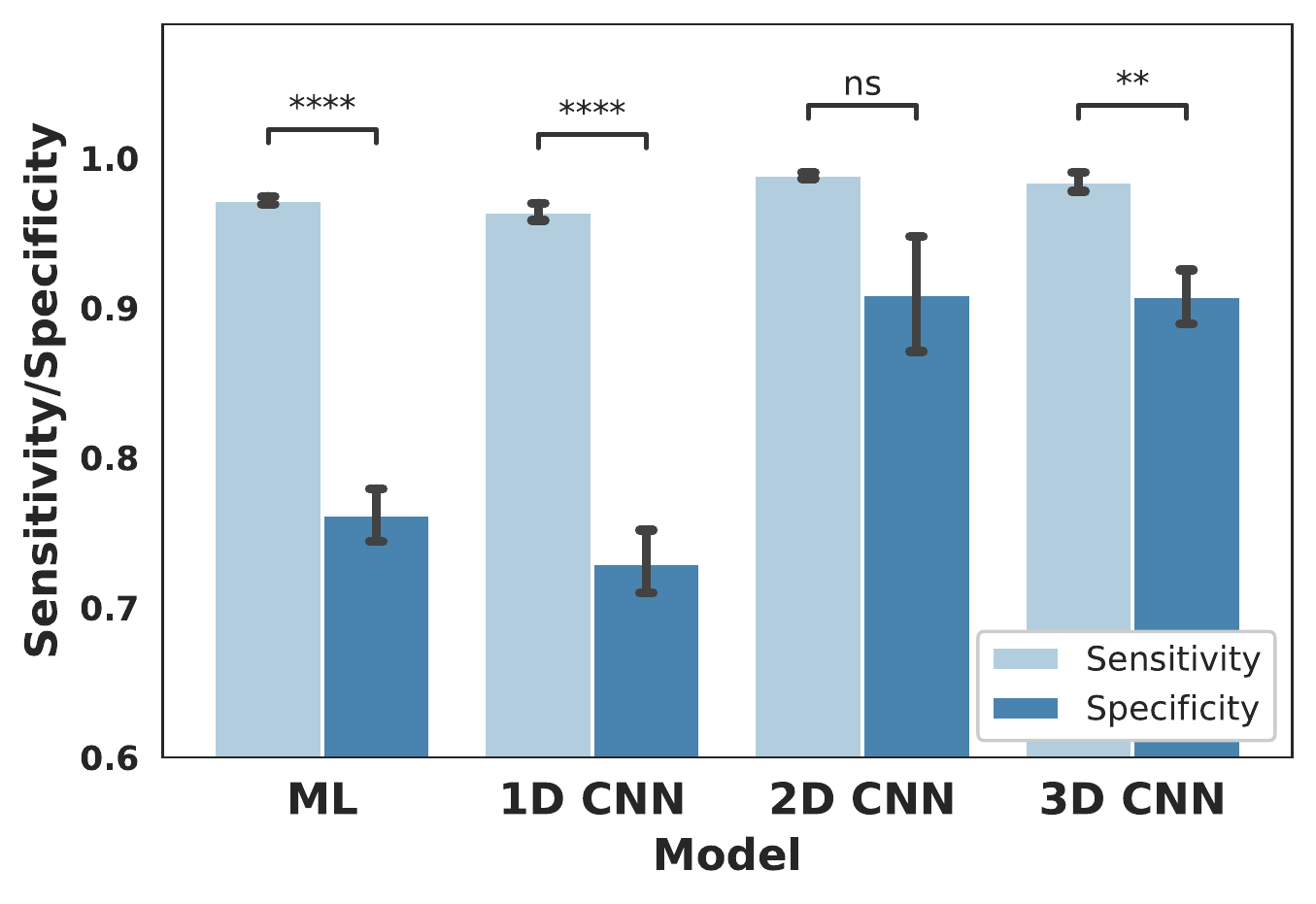}
  %\caption{A subfigure}
  %\label{fig:sub1}
\end{subfigure}%
%\hspace{-2mm}
\begin{subfigure}
  \centering
  \includegraphics[width=.32\linewidth]{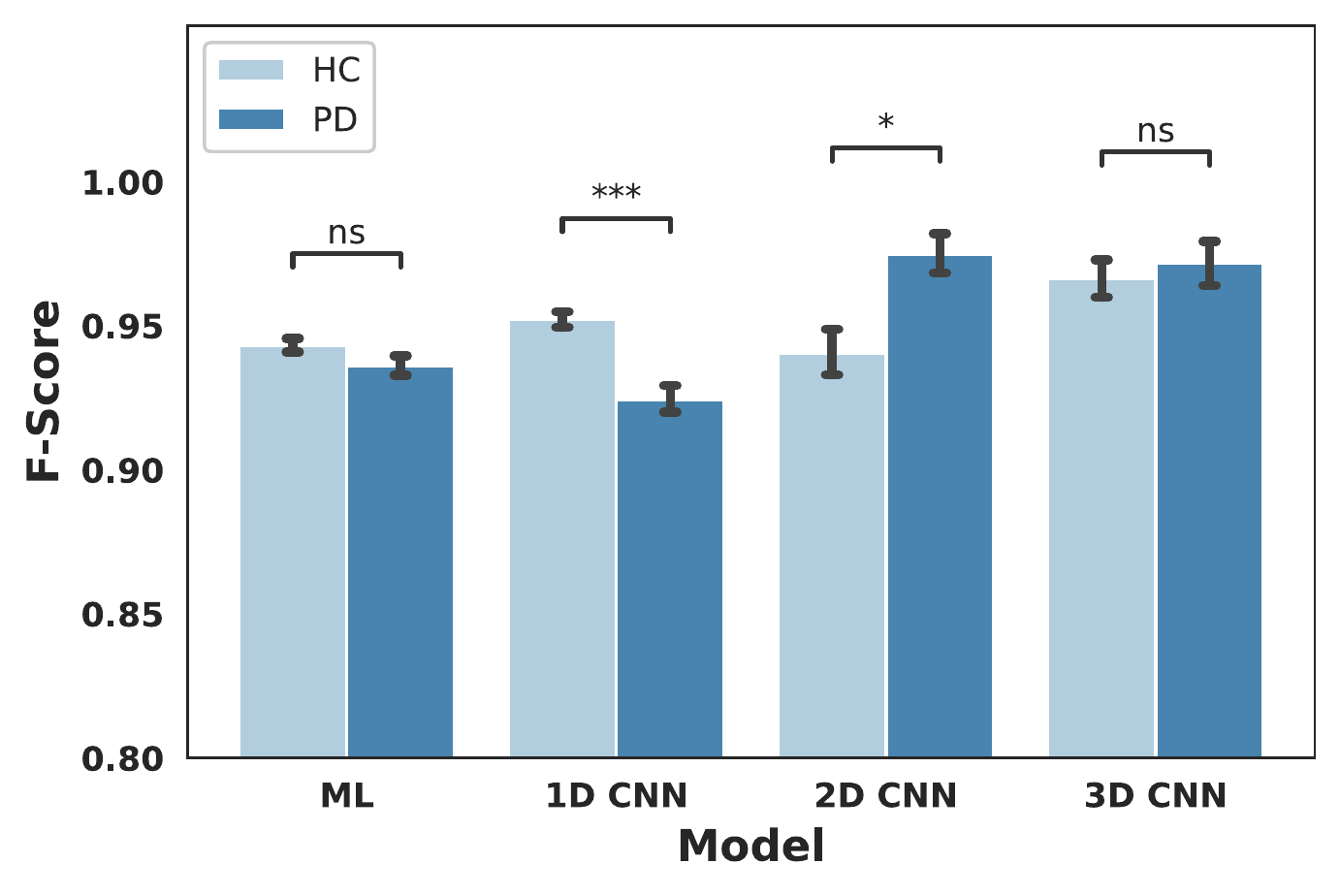}
  %\caption{A subfigure}
  %\label{fig:sub2}
\end{subfigure}
\vspace{-3mm}
\caption{\textbf{Binary arousal classification:} (Left) Sensitivity and specificity with PD data across models. (Right) F1 on PD and HC data across models. Error bars denote SEM. ****, ***, **, * and n.s. respectively imply $p < 0.0001, 0.001, 0.01, 0.05$ and $p > 0.05$ as per a Tukey HSD test.} 
\label{fig:HALA_fig}\vspace{-2mm}
\end{figure*}

\subsection{Arousal Classification}\label{asl_results}
To examine arousal perception in PD, we grouped the \emph{anger}, \emph{disgust}, \emph{fear}, \emph{happiness} and \emph{surprise} data in the high arousal (HA) category, with the \emph{sadness} samples constituting the low arousal (LA) category as in the circumplex model~\cite{russell1980circumplex}. The HA:LA class ratio within PD, HC and full data is, thus, 5:1. 

% Table \ref{tab:HALA_Table} shows the results obtained in the binary classification of HA and LA emotions with full data, PD data and HC data. With  falling in the HA category (positive class) and \emph{sad} in the LA category (negative class), the input is in the ratio of 5:1 for HA and LA classification. 

%
%----------------------------------------
%
\subsubsection{Results}
Table~\ref{tab:HALA_Table} presents arousal classification results with full, PD and HC data. Evidently, similar F1-scores were achieved for these subsets. We again compare PD vs HC results. 

%
%------------------------------
%
\paragraph{Classification using PD Data}
Impact of features on ML classification performance is confirmed by a one-way ANOVA $(F(2,27) = 1332.94, p < 0.0001)$. CSP features achieved optimal arousal prediction and significant F1-score differences were noted via a Tukey test for CSP vs SPV $(p < 0.005)$,  CSP vs Raw $(p < 0.001)$ and SPV vs Raw $(p < 0.001)$. Higher F1-scores were obtained for the 1D-CNN with spectral features performing best, even if the differences among descriptors were not significant. The 2D and 3D-CNN models achieved an identical, near-ceiling F1 of $0.98$.

Fig.~\ref{fig:HALA_fig} (left) presents specificity (LA classification rates) and sensitivity (HA classification rate) scores for PD data. Significantly higher sensitivity ($0.97$) than specificity ($0.76$) was observed for ML algorithms ($p < 0.0001)$. This trend repeated for the 1D-CNN ($p < 0.0001$) and 3D-CNN (sensitivity $=0.98>$ specificity $=0.91$ with $p < 0.01$), while comparable measures were achieved for the 2D-CNN. Overall, higher sensitivity than specificity was achieved on PD data with the different models.

% . Whereas in 3D CNN, a higher sensitivity ($0.98 \pm 0.02$) than specificity ($0.91 \pm 0.06$) is noted with significant difference between the two $(t(18) = 3.59, p = 0.008)$. The trends confirm a higher sensitivity than specificity in the arousal classification with PD data across various models. 

%
%------------------------------
%
\paragraph{Classification using HC Data}
CSP features produced a maximum F1-score of $0.93$ with ML methods on HC data. The 1D-CNN achieved a higher F1 of 0.95 with CSP features, but all features performed comparably. F1-scores of $0.94$ and $0.97$ were achieved with the 2D and 3D-CNN, respectively, revealing that the spectral EEG image and movie descriptors effectively encode emotion information. 

%
%------------------------------
%
\paragraph{PD vs HC F1-score Comparison}
F1-scores achieved with PD and HC data are presented in Fig.~\ref{fig:HALA_fig} (right), with CSP results shown for the ML and 1D-CNN methods. Very similar F1-scores were found on PD and HC data for ML algorithms. The 1D-CNN achieved a much higher score with HC data $(p < 0.0001)$, while the trend reversed for the 2D-CNN (PD F1$=0.98 >$ HC F1$=0.94$, with $p < 0.05$). Similar F1-scores with PD and HC data were again noted with the 3D-CNN. 
\begin{table*}[!ht]
\centering
\fontsize{7}{7}\selectfont
\caption{\textbf{Categorical Emotion Classification summary:} Mean F1-scores over all emotion classes (equivalent to accuracies for balanced dataset) are of the form $\mu \pm \sigma$. Best ML results were obtained with the kNN or NB algorithms. NB results are denoted using a * symbol.}\vspace{-1mm}
\resizebox{2\columnwidth}{!}{%
\begin{tabular}{c|ccc|ccc|c|c}
%\hline
\multirow{2}{*}{\textbf{Data}} &
  \multicolumn{3}{c|}{\textbf{ML}} &
  \multicolumn{3}{c|}{\textbf{1D CNN}} &
  \multirow{2}{*}{\textbf{2D CNN}} &
  \multirow{2}{*}{\textbf{3D CNN}} \\ \cline{2-7}
     & \textbf{SPV} & \textbf{CSP}   & \textbf{Raw}     & \textbf{SPV} & \textbf{CSP}     & \textbf{Raw}     &                  &                  \\ \hline
Full & 0.64 $\pm$ 0.01    & 0.72 $\pm$ 0.01 & 0.18 $\pm$ 0.01 & 0.76 $\pm$ 0.10  & 0.69 $\pm$ 0.02 & 0.77 $\pm$ 0.22 & 0.82 $\pm$ 0.09 & 0.83 $\pm$ 0.09 \\
PD   & 0.62 $\pm$ 0.02    & 0.76 $\pm$ 0.01 & \ 0.17 $\pm$ 0.01*  & 0.77 $\pm$ 0.08  & 0.80 $\pm$ 0.05 & 0.81 $\pm$ 0.21 & 0.84 $\pm$ 0.10 & 0.88 $\pm$ 0.09 \\
HC   & 0.68 $\pm$ 0.02    & 0.74 $\pm$ 0.02 & 0.20 $\pm$ 0.01  & 0.76 $\pm$ 0.09  & 0.76 $\pm$ 0.03 & 0.78 $\pm$ 0.22 & 0.86 $\pm$ 0.08 & \textbf{0.90 $\pm$ 0.07} \\ \hline
\end{tabular}%
}
\hspace{5mm}
\label{tab:Multi_Table}\vspace{-3mm}
\end{table*}

\begin{figure*}[!t]
%\captionsetup[subfigure]{aboveskip=-1pt,belowskip=-1pt}
\centering
\savebox{\imagebox}{\includegraphics[width=.58\linewidth]{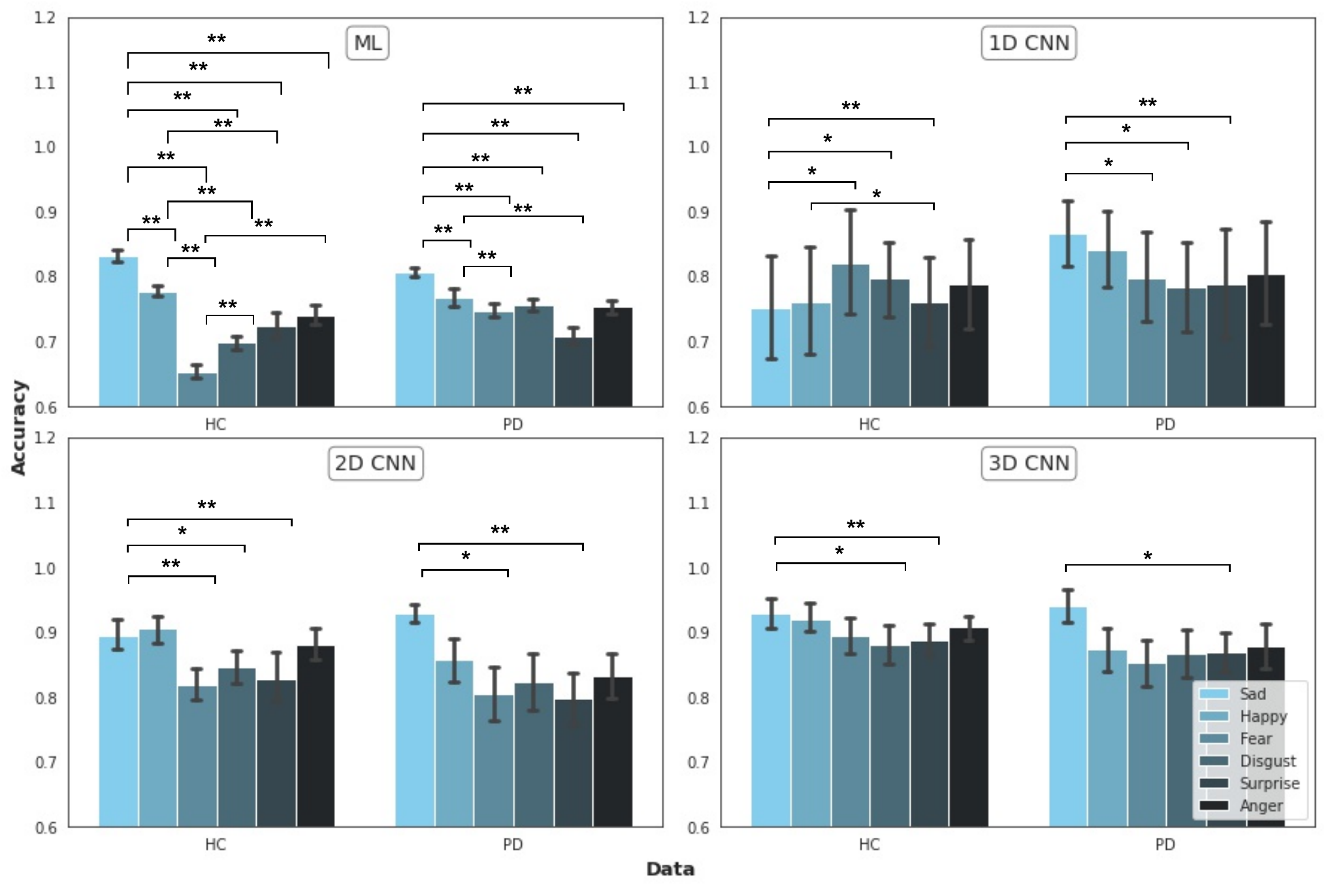}}%
  \begin{subfigure}
    \centering\usebox{\imagebox}% Place largest image
    %\caption{This is a sub-caption. This is a sub-caption. This is a sub-caption}
  \end{subfigure}
  \hspace{-1.8mm}
  \begin{subfigure}
    \centering\raisebox{\dimexpr0.5\ht\imagebox-0.48\height}{% Raise smaller image into place
      \includegraphics[width=.4\linewidth]{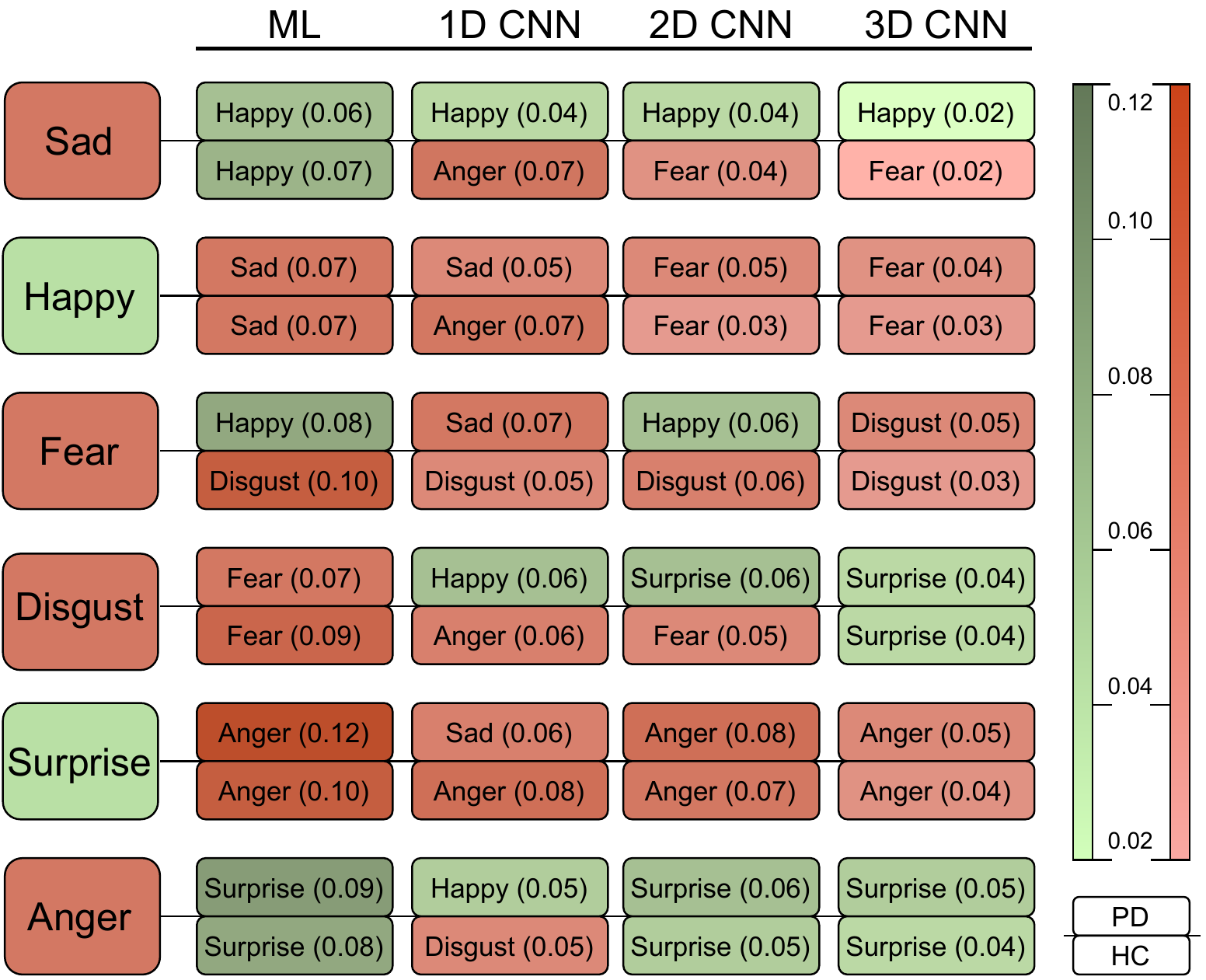}}%
    %\caption{This is a sub-caption.}
  \end{subfigure}
\vspace{-4mm}
\caption{\textbf{Categorical emotion classification:} (left) Emotion-wise F1-scores on HC (left) and PD data (right) across  models. Error bars denote the standard error of mean. ****, ***, **, * and n.s. respectively imply $p < 0.0001, p < 0.001,  p < 0.01, p < 0.05$ and $p > 0.05$ as per a Tukey HSD test. (right) Misclassifications on PD and HC emotion data across models. First column denotes the actual emotion, while other columns denote predicted emotion with the model specified on top. Upper and lower row pairs respectively denote emotions predicted with PD and HC data, and values in parenthesis denote misclassification rates. Green and red colours are used to code positive and negative valence emotions.}
\label{fig:Multi_fig}\vspace{-4mm}
\end{figure*}

\subsection{Categorical Emotion Classification} 
\label{multi_results}
{Since reduced HV-LV discriminability was noted in PD patients, we further explored  categorical emotion recognition and nature of misclassificaions with PD vs HC data.}
% explored categorical emotion recognition to examine if any (mis)classification patterns were discernible for PD patients.    

%
%------------------------------
%
\subsubsection{Results}
{Table~\ref{tab:Multi_Table} presents multi-class emotion classification results across models with full, PD and HC data. An equal number of samples were available for the \emph{sadness}, \emph{happiness}, \emph{fear}, \emph{disgust}, \emph{surprise} and \emph{anger} emotion classes for both PD and HC subjects. F1-scores averaged over all emotion classes are shown. For most conditions, highest scores achieved with HC data, while lowest scores were achieved with full data.} %PD and HC data results are discussed below. 

%
%--------------------
%
\paragraph{Classification with PD Data}\label{multi_PD}
As with valence and arousal, EEG features significantly impacted ML performance as per a one-way ANOVA $(F(2,27) = 4113.91, p < 0.0001)$. CSP features produced the best F1-score of 0.76, significantly outperforming SPV and Raw features $(p < 0.001)$ as per post-hoc Tukey tests. For the 1D-CNN, raw EEG achieved the best F1-score of $0.81$, which was marginally superior to CSP and SPV features. Higher mean F1-scores of $0.84$ and $0.88$ were achieved with the 2D and 3D-CNN, conveying accurate emotion recognition with PD data.

Fig.~\ref{fig:Multi_fig} (left) depicts emotion-specific F1-scores obtained on the PD and HC data across models, with CSP results presented for the ML and 1D-CNN models. On PD data, ML methods produced the highest and lowest F1-scores for sadness (F1$=0.81$) and surprise (F1$=0.71$), respectively, and a significant variation in F1-scores for different emotions was found per one-way ANOVA $(F(5,54) = 8.68, p < 0.0001)$. A significant effect of emotions on F1-scores was also noted for 1D, 2D and 3D-CNN ($p < 0.05$ in all cases). Sadness was easiest to recognise with all three models (F1$=0.87, 0.93$ and $0.94$ for 1D, 2D and 3D-CNN), while disgust (F1$=0.78$), surprise (F1$=0.80$) and fear (F1$=0.85$) were recognised worst by the 1D, 2D and 3D-CNN, respectively. Across models, sadness was easiest to recognise, while disgust, fear and surprise were commonly confused with other emotions. 

%
%--------------------
%
\paragraph{Classification with HC Data}
Trends similar to PD were observed with HC data. CSP features produced the highest score (F1$=0.74$) with ML methods, significantly outperforming SPV and Raw features ($p < 0.001$ for both comparisons). All features performed comparably with the 1D-CNN, while the EEG image and movie descriptors produced mean F1-scores of 0.86 and 0.90 respectively via the 2D and 3D-CNN frameworks.

%  In 3D CNN, a one-way ANOVA reveals a marginal effect of features on the emotions $(F(5,114) = 2.36, p = 0.04)$, and a \emph{post hoc} Tukey HSD test shows a significant difference between sad and fear, and sad and surprise. Sad ($0.94 \pm 0.08$) has the highest classification accuracy and fear ($0.85 \pm 0.12$) has the least accuracy. 

%
%--------------------
%
\paragraph{Misclassification Analyses}
{Hitherto, (a) sensitivity-specificity analyses show lower recognition rates for HV emotions, and (b) Emotion-specific results convey that surprise, disgust and fear are often confounded with other emotions. We further examined the nature of misclassifications with PD and HC data to discover any underlying patterns.}

Fig.~\ref{fig:Multi_fig} (right) depicts the maximum misclassification rate and most mis-predicted label per model and emotion class. For instance, the first row shows that the sad PD samples are often misclassified as happy by the best-performing ML, 1D, 2D and 3D-CNN models, with the misclassification rates specified in brackets. For HC data (2nd row), sadness is respectively mislabeled as happy, anger and fear. We note that:
\begin{itemize}
    \item The happiness and surprise high-valence emotions are most commonly mislabeled as low-valence emotions, namely, sadness, fear and anger for both PD and HC data. As per Fig.~\ref{fig:HVLV_fig} (right), misclassification rates are slightly higher with PD data than HC data. 
    \item  Among low-valence emotions, sadness is consistently predicted as happiness with PD data. Conversely on HC data, sadness is often confounded with other low-valence emotions such as fear and anger.
    \item Fear and disgust are often misclassified with both PD and HC data. On HC data, fear is frequently confounded with disgust, and disgust with fear and anger. With PD data, however, disgust is often confused with happiness and surprise, and fear with happiness. 
    \item From the above, one can infer a greater propensity to confound with opposite-valence emotions on PD data. Overall, our findings convey that valence-related differences are not effectively encoded in PD EEG responses. 
\end{itemize}

\begin{table*}[!t]
\centering
\caption{\textbf{PD vs HC classification:} F1-scores (equivalent to accuracy scores for balanced dataset) are of the form of $\mu \pm \sigma$. Best ML results are achieved with the kNN and NB classifiers. NB results are denoted with a * symbol.}\vspace{-2mm}
\begin{tabular}{c|ccc|ccc|c}
%\hline
\multirow{2}{*}{\textbf{Data}} & \multicolumn{3}{c|}{\textbf{ML}}                    & \multicolumn{3}{c|}{\textbf{1D-CNN}}                & \multirow{2}{*}{\textbf{2D-CNN}} \\ \cline{2-7}
                               & \textbf{SPV} & \textbf{CSP}   & \textbf{Raw}   & \textbf{SPV} & \textbf{CSP}   & \textbf{Raw}   &                                  \\ \hline
Full                           & 0.97 $\pm$ 0.01    & 0.88 $\pm$ 0.01 & 0.66 $\pm$ 0.01 & \textbf{0.99 $\pm$ 0.01}    & 0.87 $\pm$ 0.02 & 0.91 $\pm$ 0.07 & \textbf{0.99 $\pm$ 0.01}                   \\
Sadness                        & 0.97 $\pm$ 0.01    & 0.92 $\pm$ 0.01 & 0.58 $\pm$ 0.03 & 0.97 $\pm$ 0.02    & 0.94 $\pm$ 0.02 & 0.96 $\pm$ 0.06 & \textbf{0.99 $\pm$ 0.01}                   \\
Happiness                      & 0.97 $\pm$ 0.01    & 0.90 $\pm$ 0.02 & 0.59 $\pm$ 0.02* & 0.91 $\pm$ 0.08    & 0.92 $\pm$ 0.02 & 0.95 $\pm$ 0.09 & 0.98 $\pm$ 0.02                   \\
Fear                           & 0.96 $\pm$ 0.01     & 0.93 $\pm$ 0.02 & 0.61 $\pm$ 0.03* & 0.96 $\pm$ 0.02    & 0.95 $\pm$ 0.02 & 0.89 $\pm$ 0.15 & \textbf{0.99 $\pm$ 0.01}                   \\
Disgust                        & 0.96 $\pm$ 0.02     & 0.91 $\pm$ 0.02 & 0.59 $\pm$ 0.01* & 0.98 $\pm$ 0.03    & 0.93 $\pm$ 0.03 & 0.93 $\pm$ 0.09 & 0.98 $\pm$ 0.02                   \\
Surprise                       & 0.95 $\pm$ 0.01    & 0.91 $\pm$ 0.03 & 0.60 $\pm$ 0.02* & 0.92 $\pm$ 0.03    & 0.94 $\pm$ 0.02 & 0.94 $\pm$ 0.10 & 0.97 $\pm$ 0.03                   \\
Anger                          & 0.96 $\pm$ 0.01    & 0.92 $\pm$ 0.01  & 0.59 $\pm$ 0.03* & 0.98 $\pm$ 0.01    & 0.94 $\pm$ 0.03 & 0.87 $\pm$ 0.14 & 0.98 $\pm$ 0.02                   \\ \hline
\end{tabular}

\hspace{5mm}
\label{tab:PDNC_Table}\vspace{-4mm}
\end{table*}

\begin{figure*}[!ht]
%\captionsetup[subfigure]{aboveskip=-1pt,belowskip=-1pt}
\centering
\begin{subfigure}
  \centering
  \includegraphics[width=.32\linewidth]{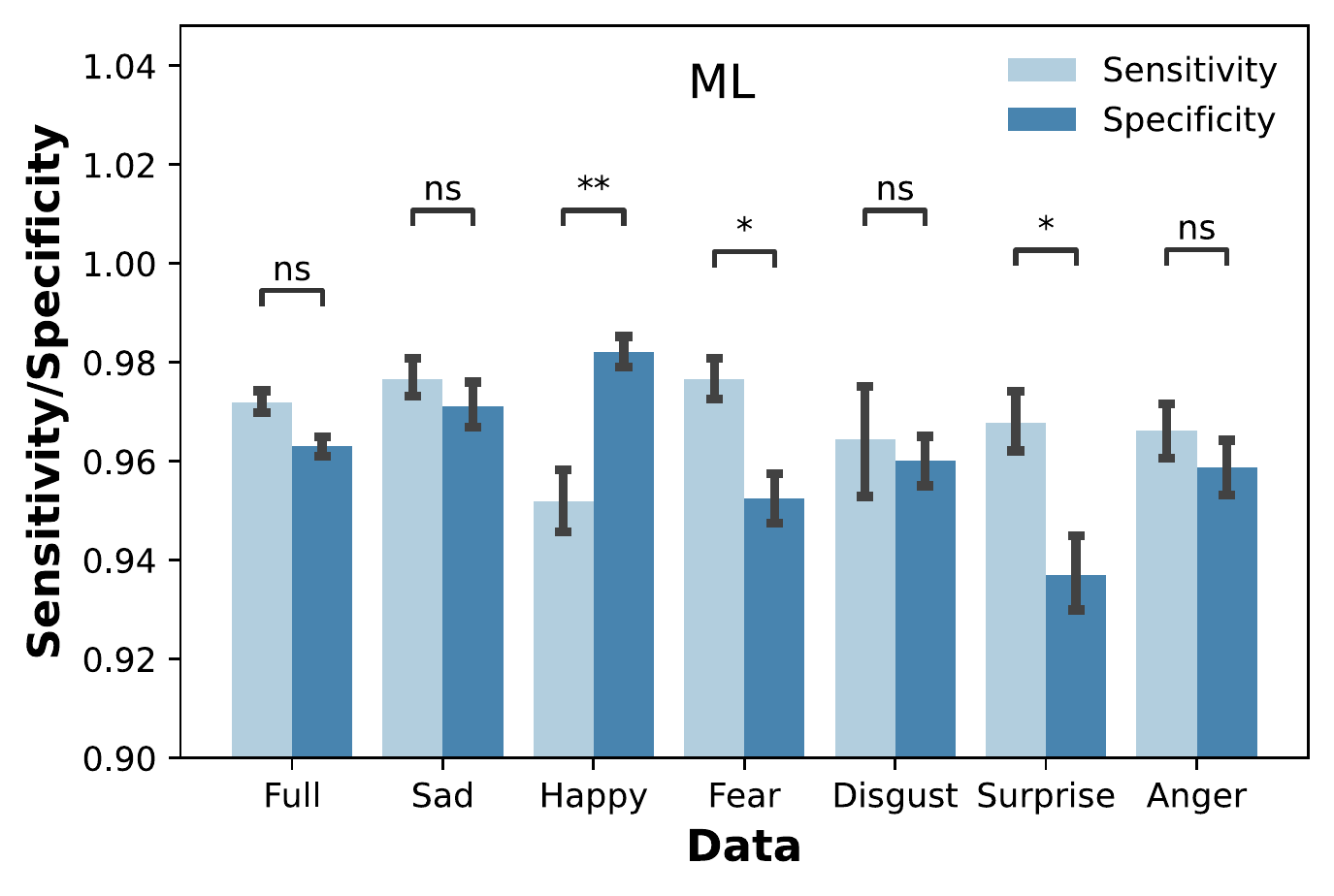}
  %\caption{A subfigure}
  %\label{fig:sub1}
\end{subfigure}%
%\hspace{-2mm}
\begin{subfigure}
  \centering
  \includegraphics[width=.32\linewidth]{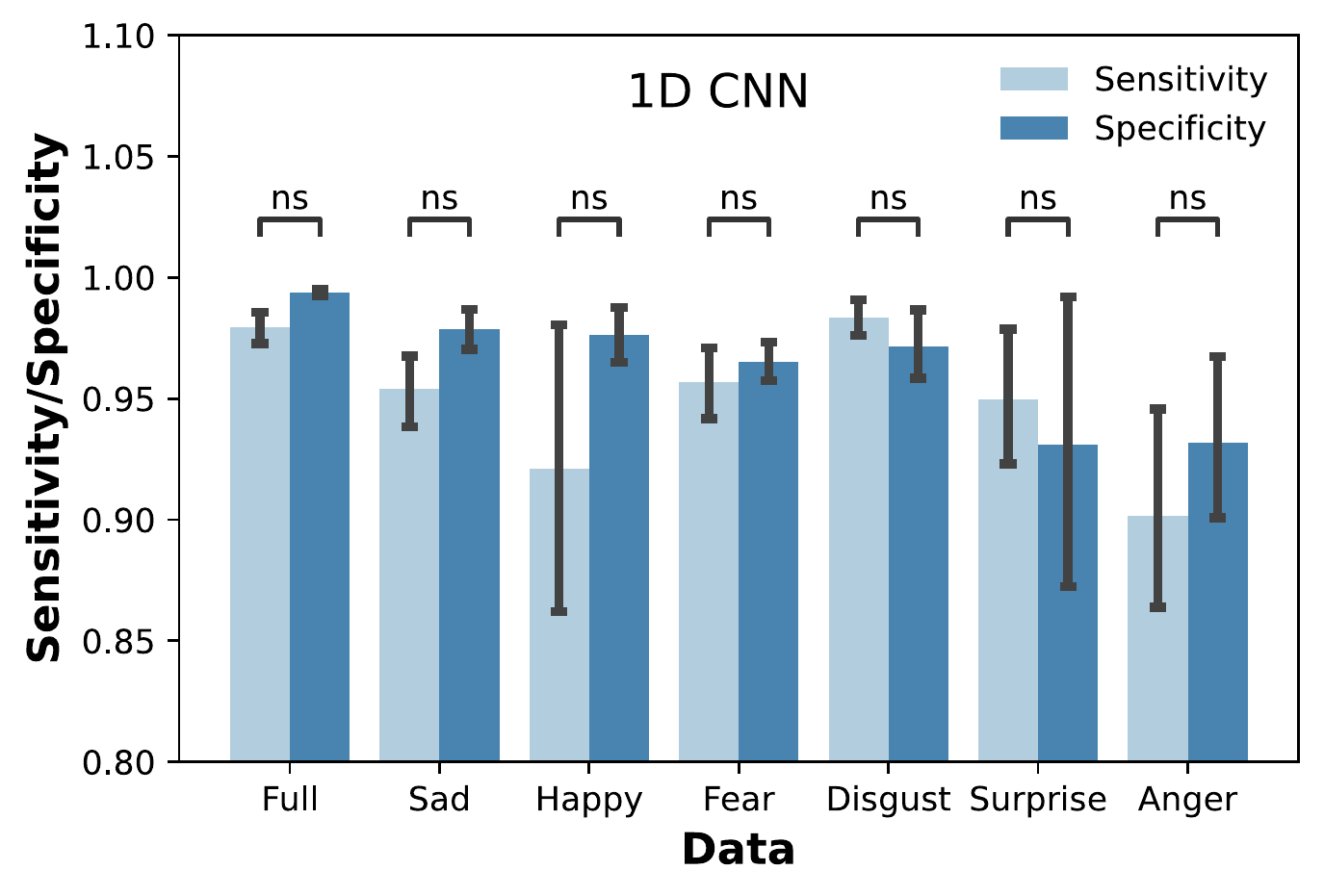}
  %\caption{A subfigure}
  %\label{fig:sub2}
\end{subfigure}
\begin{subfigure}
  \centering
  \includegraphics[width=.32\linewidth]{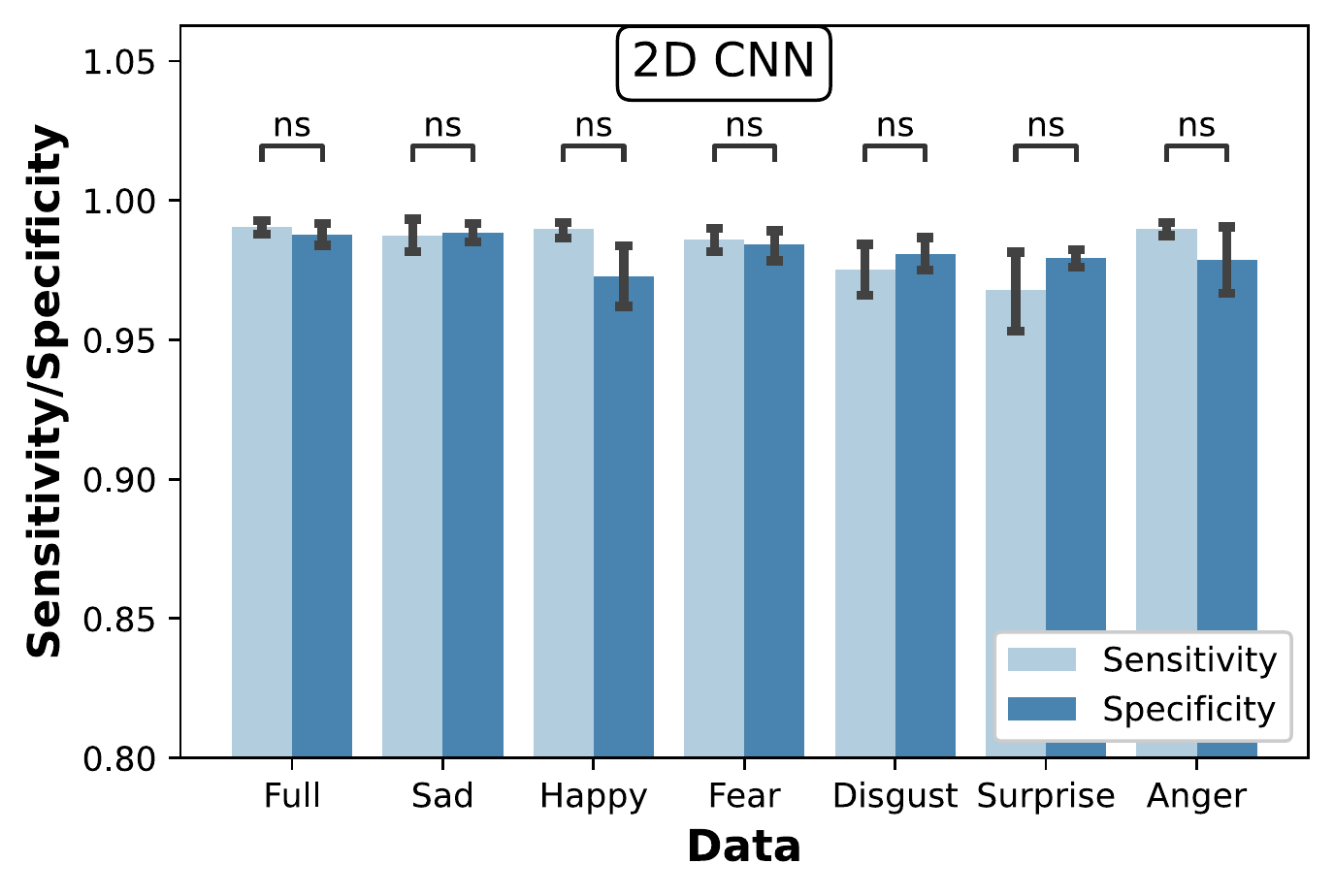}
  %\caption{A subfigure}
  %\label{fig:sub2}
\end{subfigure}
\vspace{-4mm}
\caption{\textbf{PD vs HC classification:} Sensitivity and specificity on all and emotional data with (left) ML algorithms, (centre) 1D-CNN and (right) 2D CNN. Error bars denote standard error. ****, ***, ** and * respectively $\implies p < 0.0001, 0.001, 0.01, 0.05$ as per a Tukey HSD test.}
\label{fig:PDNC_fig}\vspace{-4mm}
\end{figure*}

\subsection{PD vs HC Classification} 
\label{pdnc_results}
The above sections reveal some differences in the emotional  EEG characteristics of the PD and HC groups. We examined whether the emotional EEG responses were discriminable for PD vs HC classification. To this end, we attempted classification from  emotion-agnostic (\textit{Full}) and emotion-specific data. Results are shown in Table~\ref{tab:PDNC_Table}. Our data included equal proportions of the PD and HC classes (1:1 class ratio). 

%PD and HC were respectively considered the positive and negative classes, with a positive-to-negative class ratio of . 

%
%------------------------------
%
\subsubsection{Results} Emotion-agnostic and specific PD recognition results are presented below.
%
%--------------------
%
\paragraph{Classification using Full Data}
While raw EEG features achieved fair recognition with ML methods (max F1$=0.66$), CSP and SPV features performed much better producing the maximum F1 of $0.97$. A one-way ANOVA revealed significant differences in feature performance  $(p < 0.0001)$. All features performed better with the 1D-CNN, with SPV features producing a near-ceiling F1$=0.99$ and outperforming the CSP and Raw features as per a post-hoc Tukey test ($p < 0.001$ for both comparisons). 2D-CNN performed marginally better, achieving a mean F1$=0.99$ on full, sadness and fear data. 
 
Fig.~\ref{fig:PDNC_fig} presents sensitivity and specificity rates on full and emotion-specific data with different methods. ML algorithms on full data achieved a marginally higher sensitivity ($0.97$) than specificity ($0.96$); with the 1D CNN, the specificity ($0.99$) was marginally higher than sensitivity ($0.98$). Identical sensitivity and specificity rates of $0.99$ were noted on full data with the 2D CNN. 

%
%--------------------
%
\paragraph{Classification with emotion-specific Data}
Facile PD vs HC classification was achieved with all emotion classes, where only ${1}$/${6}^{th}$ of the full data were available. Raw EEG and SPV features respectively produced the lowest and highest F1-scores with ML methods,  with SPV F1$=0.97$ for happiness and sadness. Mixed trends were noted for the 1D-CNN, where all features performed comparably. Maximum F1 for the raw, CSP and SPV descriptors were respectively noted for sadness (F1$=0.96$), fear (F1$=0.95$) and anger (F1$=0.98$). Peak and near-ceiling classification performance was noted for the 2D-CNN (identical F1$=0.99$ for sadness and fear).

Similar sensitivity and specificity rates were noted across classifiers with emotion-specific data (see Fig.~\ref{fig:PDNC_fig}). For the 1D-CNN, near-identical sensitivity and specificity rates were noted for fear (sensitivity = 0.96 vs specificity = 0.97) and disgust (sensitivity = 0.98 vs specificity = 0.97), while for the 2D-CNN, near-identical sensitivity and specificity were noted for all emotion categories.

\section{Discussion}
\subsection{Valence \& Arousal Classification}
\subsubsection{Valence} We examined PD valence perception, since valence is a fundamental emotional attribute~\cite{sander2009oxford,shuman2013levels}. While prior studies have found valence-related differences between PD and HC groups via their explicit responses to visual~\cite{baggio2012structural}, verbal~\cite{jin2017altered} and textual~\cite{kan2002recognition} stimuli, we differently examined \emph{implicit} emotional EEG responses to this end. Valence classification with multiple features and methods convey reduced performance and lower sensitivity on PD data. Therefore, PD data exhibits lower valence discriminability and sensitivity.

Lower PD sensitivity is consistent with findings in~\cite{lin2016degraded}, where PD patients are found to have deficits in processing both positive and negative emotions. Dysfunction of the basal ganglia thalamocortical circuits in PD patients impairs their general emotional valence recognition~\cite{pillay2007recognition}. 
With respect to models and features, CNNs expectedly achieved higher F1-scores than ML methods, confirming that they can efficiently learn spatio-temporal EEG patterns~\cite{yue2015beyond}. CSP features predominantly achieve the best scores, and their utility in EEG-based analysis is well known~\cite{ramoser2000optimal}. The EEG-movie descriptor optimally encodes spatio-temporal patterns in spectral EEG.

\subsubsection{Arousal}
{With respect to arousal,~\cite{miller2009startle, bowers2006startling} observed muted reactivity or fewer startled eye-blinks from PD patients to high-arousal, low-valence aversive pictures. Similar findings were reported in~\cite{kan2002recognition} and~\cite{paulmann2010dynamic}, where PD patients showed deficits in recognizing emotions from lexical, prosody and facial cues. While these findings are based on implicit EMG data and/or explicit self-ratings, our inferences are based on EEG classification patterns. {Given the identical class distributions for PD and HC data and corresponding F1 comparisons, our results do not indicate any PD vs HC differences in arousal recognition. Very comparable F1-scores are obtained for the PD and HC groups, conveying their similar characteristics.% for the PD and HC groups. %While high F1-scores are noted {for} both groups for valence classification, slightly higher scores are consistently noted for the PD group with all features and models. This trend suggests lower HV vs LV discriminability with PD EEG data.

\subsubsection{Models \& Features} Sensitivity and specificity scores observed for PD data also convey an interesting trend. Given the imbalanced class proportions for both the valence and arousal conditions, significant sensitivity vs specificity disparities are noted particularly for the ML and 1D-CNN models. However, these differences become less conspicuous for the 2D and 3D-CNN models, conveying that they are able to efficiently learn minority-class representations. Regarding algorithms and features, F1-scores gradually improve while advancing from ML algorithms to the 3D-CNN. CSP features performed best with ML algorithms, but mixed results were observed for the 1D-CNN. EEG image and movie descriptors achieved a maximum and identical F1-score of $0.98$ for arousal, with the 3D-CNN producing the maximum F1 of $0.93$ for valence, showing their efficacy in encoding emotional information.}

\subsection{Categorical Emotion Recognition}
To the best of our knowledge, we are the first to examine categorical emotional classification with PD and HC EEG data. Prior studies on PD emotional perception typically examined facial expression recognition tasks~\cite{narme2011understanding,clark2008specific,kan2002recognition} or studied physiological signals along with self-assessment reports to understand emotional deficits \cite{wabnegger2015facial, wu2014objectifying, dietz2011emotion}. These studies observe PD impairment in recognising negative emotions such as \emph{sadness}, \emph{fear}, \emph{anger} and \emph{disgust}.

We performed categorical emotion recognition to better understand which emotions are recognized better/worse with PD and HC data. Our results revealed that while sadness and happiness were well recognized with both PD and HC groups, fear, disgust and surprise were poorly recognized with the PD EEG data. Disturbances in the orbitofrontal cortex and the anterior cingulate cortex, which are active in negative emotion processing, can be attributed to these deficits~\cite{lin2016degraded}.
We then studied the nature of misclassifications for each emotion class. Classification results in Fig.\ref{fig:Multi_fig} (right) show frequent confounds among opposite-valence emotions with PD data, indicating weaker valence encodings in emotional PD responses.

With respect to features and models, the trends are consistent with valence and arousal classification. We observe a steady increase in emotion-specific and overall F1-scores as we progress from classical machine learning methods to the 3D-CNN. CSP and raw EEG features produce the best performance with the ML and 1D-CNN approaches respectively. The 2D and 3D-CNN models, however, achieve higher F1-scores, implying that spatio-temporal spectral EEG patterns best encode emotional information.

\subsection{PD vs NC classification}
{To our knowledge, only one study~\cite{YUVARAJ2014108} has performed PD recognition from emotional EEG signals, and achieved a mean accuracy of 87.9\% (see Table~\ref{rw_comp}). 
Others~\cite{bhurane2019diagnosis,Shu2020} have performed PD vs HC classification from resting-state EEG signals, which is an ecologically invalid setting requiring a highly controlled environment for EEG acquisition. In this regard, we attempted PD recognition from both emotion-specific and emotion-agnostic (or full) EEG data acquired during the routine task of audio-visual media consumption. 
% no study has been conducted using the emotional state EEG signals of PD patients and HC. 
% The classification of the EEG signals of PD and HC will facilitate the computer-aided diagnosis of PD. In this direction, we perform the binary classification of PD and HC using the full data and the data of individual emotions. 

Empirical results presented in Table~\ref{tab:PDNC_Table} show that accurate PD recognition is achieved even with emotion-specific EEG data. Near-perfect F1s are noted with the 2D-CNN, implying that  PD and HC emotional responses are highly discriminable upon learning from only a few training samples. Some correlations can also be noted between Table~\ref{tab:PDNC_Table} and Section~\ref{multi_results}. PD recognition is {highest} for the {sadness} and fear emotions in Table~\ref{tab:PDNC_Table}, and the PD emotion recognition rates are noted to be high and low, respectively, for {sadness} and fear in Sec.~\ref{multi_PD}. These findings suggest that PD-related differences may be better encoded in EEG for negative valence emotions.    

Further examining sensitivity and specificity measures, balanced recognition of PD and HC classes across models was achieved mainly for the negative disgust, fear and sadness emotions. That negative emotions best reflect PD impairments has been observed in prior studies~\cite{baggio2012structural,narme2011understanding}. Focusing on features and models, spectral features achieved the best results with ML methods. With the 1D-CNN, superior F1-scores were achieved with all features even if no clear trends were discernible. Similar to emotion recognition, 2D-CNN again achieved optimal PD recognition, demonstrating that the EEG movie features best encode PD-related emotional differences.}  
%, and categorical emotion recognition results (Fig.~\ref{fig:Multi_fig}) generally show lower recognition rates for fear and disgust on PD data. 

% In the ML algorithms with full data, a higher sensitivity than specificity is observed, while in 1D CNN and 2D CNN, almost equal sensitivity and specificity scores are observed. Also, using 2D CNN, ceiling performance is achieved in the classification of PD and HC. These observations imply that the model is able to classify the data of PD and HC to near perfection, outperforming the accuracies in the works. Hence, such a model can be deployed for the computer-aided diagnosis of PD. Apart from full data, using 2D CNN, we achieve a ceiling performance in the classification using the data of individual emotions. The sensitivity and specificity scores in the classification with each emotion are almost equal, further strengthening our finding. 

% In the ML algorithms, SPV features have outperformed the CSP and Raw features in the classification with full data and with the data of individual emotions. While in 1D CNN, the SPV features result in a ceiling performance with full data. Our results of ceiling performance in both sensitivity and specificity highlights that our model will perform accurately when used in a computer-aided diagnosis system. 

% Overall, as in the case of the other classification tasks, we observe a similar trend of increasing performance from ML algorithms to 2D CNN in the classification of PD and HC data. 

%
%----------------------------------------
%
\section{Conclusions}
\label{conclusion}
{Parkinson's disease patients may often have difficulty expressing their emotions and internal feelings in real-life owing to (a) PD effects especially in its advanced stages, and (b) the effect of associated medications. Given these limitations, an assistive and sustainable {diagnostic} tool based on non-invasive detection of emotional disturbances can facilitate treatment and help improve life quality for PD patients. While many studies identify PD-related impairments based on the patients' explicit and implicit responses~\cite{smith1996spontaneous,paulmann2010dynamic,miller2009startle}, or cognitive dissimilarities based on resting-state EEG~\cite{Wang2020,bhurane2019diagnosis,Shu2020}, we differently examined emotional EEG responses to achieve both emotion and PD recognition.

While studies examining facial behavior and resting-state EEG~\cite{Wang2020} typically derive their findings based on statistical patterns observed for the PD and HC groups, our inferences are entirely derived from classification patterns. Interesting trends and similarities with prior work were revealed our analyses. Dimensional emotion recognition experiments conveyed reduced discriminability with PD EEG data, while arousal-related differences vis-\'a-vis the HC group were not apparent. Furthermore, categorical 
emotion recognition results revealed that disgust, fear and surprise were associated with low recognition rates on PD data, while sadness was well recognized. Mislabeling analyses showed frequent confounds among opposite-valence emotions with PD data, but not with HC data. Reduced recognition of low-valence emotions, and confounds noted with positive emotions mirrors with deficits noted in the perception of these emotions from pictorial~\cite{miller2009startle, bowers2006startling} and prosodic stimuli~\cite{paulmann2010dynamic}.

%Overall, low valence was better recognised than high valence on PD data across different models. The categorical  Misclassifications showed that among PD samples, happiness and surprise were commonly confounded with low-valence emotions, while sadness, fear and disgust were often mislabeled as happiness or surprise. The propensity for confusion with opposite valence emotions is not prominent in HC data.
Given some differences in emotion perception between the PD and HC groups, we then examined if the PD vs HC emotional responses were discriminable, and if this discriminability differed across emotions. Empirical results revealed that differences were apparent for both emotion-specific and emotion-agnostic data, with high F1-scores achieved in all conditions. Here again, the maximum F1 of 0.99 was achieved for the sad and fear emotions, which respectively corresponded to a high and low recognition rate on PD data. Also, most similar sensitivity and specificity rates across models were noted for negative emotions such as disgust, sadness and fear, implying that the PD and HC classes were most discriminable for these emotions. 

With respect to features and models, CSPs considerably outperformed  spectral features with machine learning models for emotion recognition. Conversely, spectral descriptors outperformed CSPs for PD vs HC classification. The efficacy of spectral features for isolating PD characteristics has been observed in prior studies~\cite{Neufeld94}. No single feature performed best with the 1D-CNN, even if the 1D-CNN consistently outperformed classical ML methods. The 2D and 3D-CNN models consistently achieved optimal recognition performance, conveying that spectral spatio-temporal models best encode EEG patterns as noted in~\cite{bashivan2015learning,Chugh20}.

The key finding from our presented study is that both emotion and PD recognition can be reliably performed from EEG responses passively compiled during audio-visual stimulus viewing; given that we effortlessly interact with media routinely, EEG signals can be captured easily over longer time-intervals as compared to resting-state EEG, which can practically be acquired only over short episodes. Also, while many EEG differences between PC and NC groups have been noted from rest-state analysis, the exact relation between EEG and motor symptoms is unknown~\cite{Wang2020}.

Study limitations include analysis of data compiled from a limited number of PD subjects with only mild-to-moderate disease severity (Hoehn and Yahr scale~\cite{Hoehn1998} of 1--3). Future work will also focus on severity levels 4 and 5. While perceptual differences between PD vs HC subjects were captured via classification results in this study, an assistive diagnostic tool should also be able to provide \emph{explanations} behind decision-making. Future work will focus on generating explanatory predictions, building on recent work~\cite{ijcai2019-932} in this regard.}

\section{Acknowledgement}
This research was supported partially by the Australian Government through the Australian Research Council’s Discovery Projects funding scheme (project DP190101294).

\bibliography{short.bib, references.bib}{}
\bibliographystyle{IEEEtran}

%
% <OR> manually copy in the resultant .bbl file
% set second argument of \begin to the number of references
% (used to reserve space for the reference number labels box)
%\begin{thebibliography}{1}

%\end{thebibliography}

% biography section
% 
% If you have an EPS/PDF photo (graphicx package needed) extra braces are
% needed around the contents of the optional argument to biography to prevent
% the LaTeX parser from getting confused when it sees the complicated
% \includegraphics command within an optional argument. (You could create
% your own custom macro containing the \includegraphics command to make things
% simpler here.)
%\begin{IEEEbiography}[{\includegraphics[width=1in,height=1.25in,clip,keepaspectratio]{mshell}}]{Michael Shell}
% or if you just want to reserve a space for a photo:

\vskip -3\baselineskip% plus -1fil
%\vskip -2\baselineskip plus -1fil
\begin{IEEEbiography}[{\includegraphics[width=1in,height=1.25in, clip, keepaspectratio]{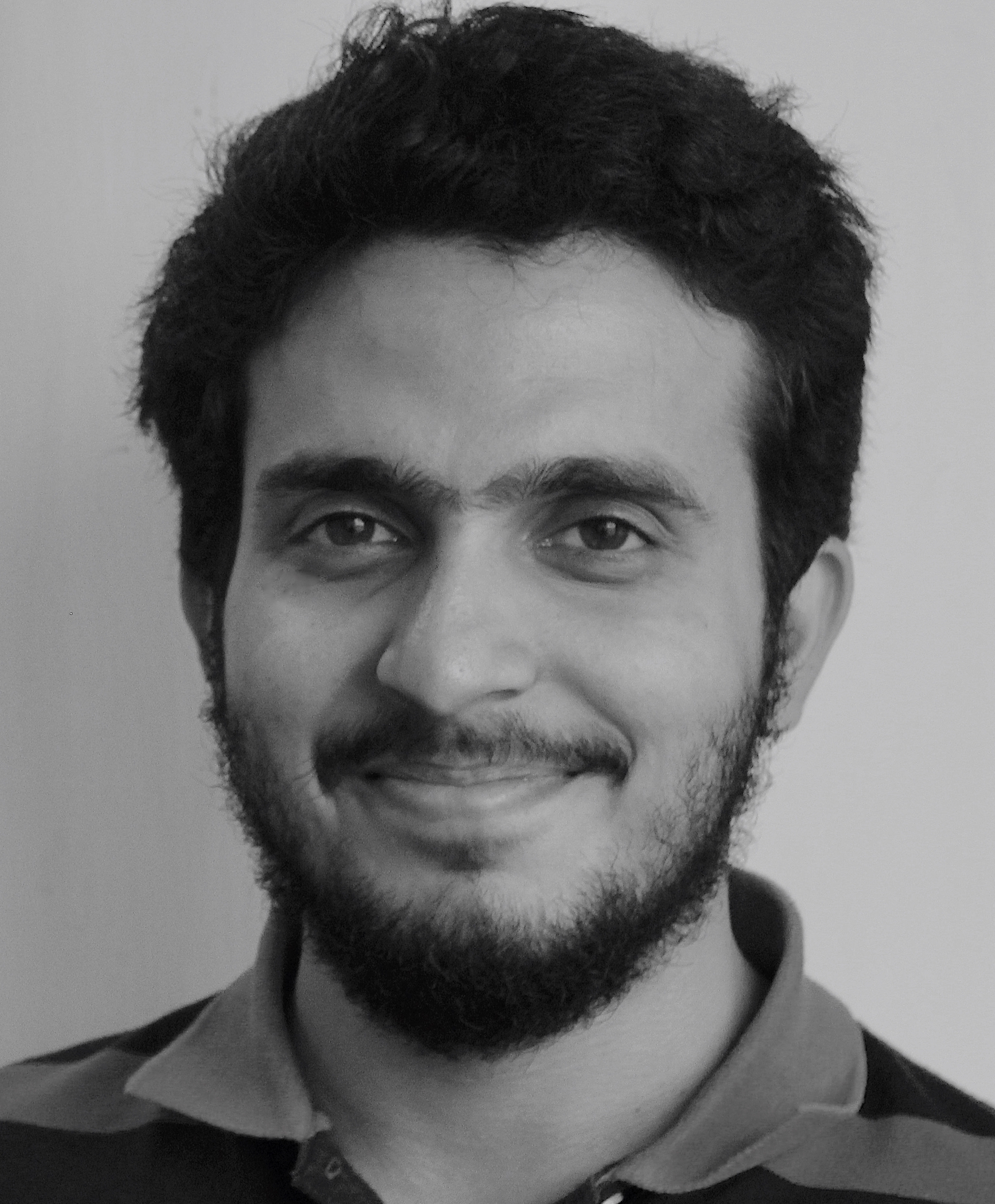}}]{Ravikiran Parameshwara} is a Ph.D.\ student at the University of Canberra, Australia. He completed his Masters in Mathematics from Christ University, Bangalore, India in 2017, and his Bachelors of Science degree from St. Joseph's college, Bangalore in 2015. His research interests broadly lie in the fields of affective computing, computer vision, and human-computer interaction. He is a student member of the IEEE.
\end{IEEEbiography}
\vskip -3\baselineskip plus -1fil
\begin{IEEEbiography}[{\includegraphics[width=1in,height=1.25in, clip, keepaspectratio]{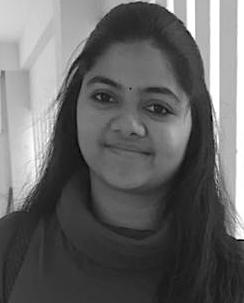}}]{Soujanya Narayana} received her Bachelor's degree from Regional Institute of Education, Mysore, India in 2015 and Masters degree from Christ University, Bengaluru, India in 2017. She is currently a PhD candidate in the Human-Centred Technology Research Centre at the University of Canberra, Australia. Her research interests include affective computing in human-computer interaction and computer vision. She is a student member of the IEEE.
\end{IEEEbiography}
\vskip -3\baselineskip plus -1fil
\begin{IEEEbiography}[{\includegraphics[width=1in,height=1.25in,keepaspectratio]{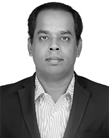}}]{M.Murugappan} received his Ph.D.\ (Mechatronic Engineering) from Universiti Malaysia Perlis, Malaysia in 2010. He is Associate Professor in the Department of Electronics and Communication Engineering, Kuwait College of Science and Technology (KCST). Previously, he was a Senior Lecturer at School of Mechatronic Engineering, Universiti Malaysia Perlis (UniMAP). He is among the top 2\% scientists in Experimental Psychology and Artificial Intelligence, and has published more than 110 research articles. 
\end{IEEEbiography}
% He secured nearly $2.5 Million as research grants for his research works from Malaysia, and Kuwait for his research works and successfully guided 14 postgraduate students (9 Ph.D. and 5 M.Sc). He is serving as an Editorial board member in PLOS ONE, Journal of Medical Imaging and Health Informatics, and International Journal of Cognitive Informatics. He is currently serving as a Chair of Educational Activities in the IEEE Kuwait Section. His main research interests are Affective Computing, Internet of Things (IoT), Brain-Computer Interface, Neuromarketing, Medical Image Processing, and Artificial Intelligence.

\vskip -2.5\baselineskip plus -1fil
\begin{IEEEbiography}[{\includegraphics[width=1in,height=1.25in,keepaspectratio]{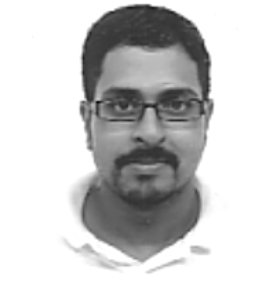}}]{Ramanathan Subramanian} received his Ph.D.\ in Electrical and Computer Engg. from NUS in 2008. He is Associate Professor in the School of IT \& Systems, University of Canberra. His past affiliations include IIT Ropar, IHPC (Singapore), U Glasgow (Singapore), IIIT Hyderabad and UIUC-ADSC (Singapore). His research focuses on Human-centered computing, especially on modeling non-verbal behavioral cues for interactive analytics. He is an IEEE Senior Member, and an ACM and AAAC member.
\end{IEEEbiography}

\vskip -2.5\baselineskip plus -1fil
\begin{IEEEbiography}[{\includegraphics[width=1in,height=1.25in,keepaspectratio]{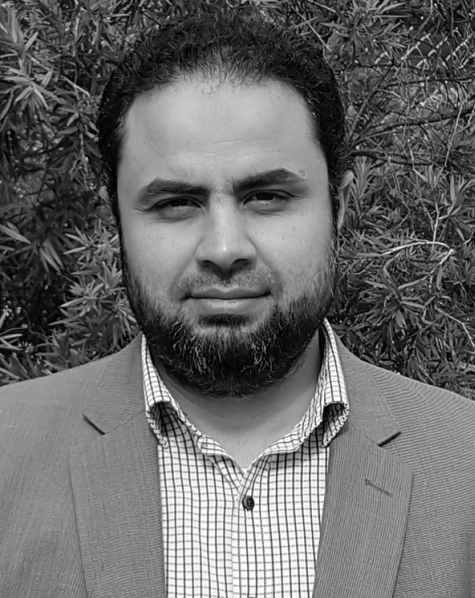}}]{Ibrahim Radwan} received the Ph.D.\ degree in computer science from the University of Canberra in 2015. From 2014 to 2016, he was a researcher in a leading automotive industry warehouse, Research Fellow with The Australian National University and currently an assistant professor at the University of Canberra. His research includes computer vision, machine learning, robotics, and artificial intelligence.
\end{IEEEbiography}

\vskip -2.5\baselineskip plus -1fil
\begin{IEEEbiography}[{\includegraphics[width=1in,height=1.25in,keepaspectratio]{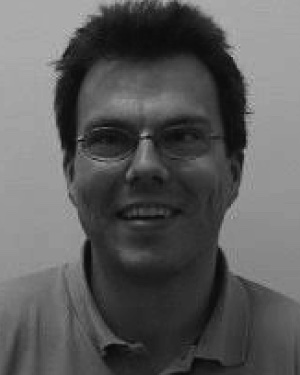}}]{Roland Goecke} received his Ph.D.\ degree in computer science from The Australian National University, Canberra, in 2004. He is Professor of Affective Computing with the University of Canberra, where he is serves as Director of the Human-Centred Technology Research Centre. His research interests include affective computing, pattern recognition, computer vision, human–computer interaction and multimodal signal processing. He is a senior member of the IEEE, and an ACM and AAAC member.
\end{IEEEbiography}

% insert where needed to balance the two columns on the last page with
% biographies
%\newpage
% if you will not have a photo at all:

% You can push biographies down or up by placing
% a \vfill before or after them. The appropriate
% use of \vfill depends on what kind of text is
% on the last page and whether or not the columns
% are being equalized.

%\vfill

% Can be used to pull up biographies so that the bottom of the last one
% is flush with the other column.
%\enlargethispage{-5in}

% that's all folks

\end{document}